\documentclass[aps,prd,showpacs,eqsecnum,twocolumn,superscriptaddress]{revtex4}

\usepackage{latexsym} \usepackage{amssymb} \usepackage{amsfonts}
\usepackage{amsmath} \usepackage{bm} \usepackage[dvips]{graphicx}
\usepackage{color}
\usepackage{subfigure} \usepackage{times} \usepackage{units}
\usepackage{hyperref}
\usepackage{lipsum}
\usepackage{float}
\usepackage{latexsym}
\usepackage{envmath}

\usepackage[utf8x]{inputenc} \usepackage{amssymb,amsmath}
\usepackage{graphicx} \usepackage[squaren]{SIunits}

\begin{document}
\title{Fully covariant and conformal formulation of the Z4 system in a
  reference-metric approach: comparison with the BSSN formulation in spherical symmetry}
\author{Nicolas \surname{Sanchis-Gual}}\affiliation{Departamento de
  Astronom\'{\i}a y Astrof\'{\i}sica, Universitat de Val\`encia,
  Dr. Moliner 50, 46100, Burjassot (Val\`encia), Spain} 

\author{Pedro J. \surname{Montero}}\affiliation{Max-Planck-Institute für Astrophysik, Karl-Schwarzschild-Str. 1, 85748, Garching bei München, Germany}

\author{Jos\'e A. \surname{Font}}\affiliation{Departamento de
  Astronom\'{\i}a y Astrof\'{\i}sica, Universitat de Val\`encia,
  Dr. Moliner 50, 46100, Burjassot (Val\`encia), Spain}   

\author{Ewald \surname{Müller}}\affiliation{Max-Planck-Institute für Astrophysik, Karl-Schwarzschild-Str. 1, 85748, Garching bei München, Germany}

\author{Thomas W. \surname{Baumgarte}}\affiliation{Department of Physics and Astronomy, Bowdoin College, Brunswick, ME 04011, USA}


\begin{abstract}

We adopt a reference-metric approach to generalize a covariant and conformal version of the Z4 system of the Einstein equations.   We refer to the resulting system as ``fully covariant and conformal", or fCCZ4 for short, since it is well suited for curvilinear as well as Cartesian coordinates.
We implement this fCCZ4 formalism in spherical polar coordinates under the assumption of spherical symmetry using a partially-implicit Runge-Kutta (PIRK) method and show that our code can evolve both vacuum and non-vacuum spacetimes without encountering instabilities.  Our method does not require regularization of the equations to handle coordinate singularities, nor does it depend on constraint-preserving outer boundary conditions, nor does it need any modifications of the equations for evolutions of black holes.  We perform several tests and compare the performance of the fCCZ4 system, for different choices of certain free parameters, with that of BSSN.  Confirming earlier results we find that, for an optimal choice of these parameters, and for neutron-star spacetimes, the violations of the Hamiltonian constraint can be between 1 and 3 orders of magnitude smaller in the fCCZ4 system than in the BSSN formulation.   For black-hole spacetimes, on the other hand, any advantages of fCCZ4 over BSSN are less evident. 
\end{abstract}


\maketitle

\section{Introduction}
\label{section:intro}
Numerical relativity has become a field of intense activity and
considerable progress has been made during the last decade. The
possible detection of gravitational waves by the second-generation
enhanced detectors (Advanced LIGO \cite{LIGO}, Advanced VIRGO
\cite{VIRGO} and KAGRA \cite{KAGRA}) represents a major incentive for
the development of numerical simulations able to provide accurate
gravitational waveforms from astrophysical sources.
 
Many current numerical relativity codes use the so-called BSSN
formulation of Einstein equations, originally proposed by 
Nakamura {\it et.al.}~\cite{Nakamura} and subsequently modified by 
Shibata and Nakamura~\cite{Shibata} and Baumgarte and Shapiro~\cite{Baumgarte}.  The
stability properties of the BSSN formulation are a result of the ``conformal connection functions", which
are introduced as new independent variables.   In combination with certain gauge conditions -- in particular
the ``1+log" slicing condition~\cite{Bona97a} and the ``Gamma-driver condition"~\cite{Alcubierre02a} -- the BSSN 
formulation has allowed for accurate and stable simulations of strong-field spacetimes, including black holes and 
neutron stars.

Recently, other conformal and traceless decompositions of the Einstein
equations, based on the Z4 system~\cite{Bona}, have been proposed
by Bernuzzi and Hilditch~\cite{Sebastiano1}, the so-called Z4c formulation, and
also by  Alic {\it et.al.}~\cite{Alic1,Alic2}, the CCZ4 formulation. Unlike the BSSN
formulation, both the Z4c and CCZ4 systems incorporate 
the constraint damping scheme developed by Gundlach {\it et.al.}~\cite{Gundlach} that 
allows for the dynamical control of the constraint violations by means of constraint damping 
terms.  The Z4c system discards non-damping non-principal terms, breaking the
4-covariance, but allowing the evolution equations to take a form that is very
similar to BSSN.   The CCZ4 system, on the other hand, retains all damping terms and maintains the
4-covariance.  Nevertheless, the CCZ4 system as presented initially
in~\cite{Alic1} suffers from numerical instabilities that develop in black hole spacetimes
unless the 4-covariance is broken. This issue was addressed by Alic
{\it et.al.}~\cite{Alic2}, who prescribed a modification for the damping
parameter that removes the instabilities when using the fully
covariant version of the CCZ4 system in the evolution of black holes.  Both
conformal decompositions of the Z4 system have been tested extensively~\cite{Sebastiano1,Sebastiano2,Sebastiano3,Sebastiano4,Alic1,Alic2,Kastaun}.
Numerical results show that, in non-vacuum simulations, violations of 
the Hamiltonian constraint can be as much as 1 to 3 orders of magnitude 
smaller than those in the BSSN formulation.

Both the BSSN and the CCZ4 or Z4c formulations in their original form are developed under the assumption of 
Cartesian coordinates; in particular they assume that the determinant of
the conformal metric is equal to one.  In the case of the BSSN
formulation, this issue was resolved by \cite{Gourgoulhon07,Brown,Gourgoulhon12}, who
introduced a covariant formulation of the BSSN equations that is
well-suited for curvilinear coordinate systems by adopting a reference-metric 
framework~\cite{FCF}. This approach allows, for example, for implementations in spherical polar coordinates,
which is of great interest since many astrophysical phenomena 
are symmetric with respect to the rotation axis (e.g., accretion disks) or are such that spherical 
coordinates adapt better to their geometry (e.g., gravitational collapse). 

The singularities associated with curvilinear coordinate systems, however,
are a known source of numerical problems.  For instance, one problem
arises because of the presence of terms in the evolution equations 
that diverge like $1/r$ near the origin $r=0$.  Several methods have
been proposed to deal with the singular terms that appear in curvilinear coordinates.  
Cordero-Carri\'on {\it et.al.}~\cite{CC12} recently
adopted a partially implicit Runge-Kutta (PIRK) method to evolve
hyperbolic, wave-like equations in the Fully Constrained formulation
of  the Einstein equations (see \cite{FCF}).  Montero and Cordero-Carri\'on \cite{Pedro1}, 
assuming spherical symmetry, applied a second-order PIRK method to the BSSN 
equations and obtained stable numerical simulations of vacuum and
non-vacuum spacetimes without the need for a regularization algorithm
at the origin. This approach has been successfully implemented in
3D without any symmetry assumption by~\cite{Thomas1} and more
recently by~\cite{Pedro2} who reported the first successful implementation of 
relativistic hydrodynamics coupled to dynamical spacetimes in spherical polar 
coordinates with no symmetry assumptions.

The purpose of this paper is threefold.  We first generalize the covariant and conformal 
Z4 system using a reference-metric approach.  We refer to this new system as 
``fully covariant and conformal Z4" or fCCZ4 for short.  This approach allows us 
to write the evolution equations in a fully covariant form suitable for spherical polar
and other curvilinear coordinates.  Second, we implement the fCCZ4 
system in spherical polar coordinates under the assumption of spherical symmetry, and show 
that using the PIRK scheme we obtain robust and stable numerical evolutions of both 
vacuum and non-vacuum spacetimes. Third, we show that the fCCZ4 formulation with the 
PIRK scheme can handle spacetimes containing black holes without the appearance of
any instability and  without the need for the modification prescribed by Alic {\it et.al.}~\cite{Alic2}. 
Finally, we compare results obtained with the BSSN and the fCCZ4 formulations. Confirming earlier results, we find that, for certain choices of free parameters, fCCZ4 can significantly reduce constraint violations, in particular for neutron-star spacetimes.  For black-hole simulations, however, the advantages of fCCZ4 over BSSN are less evident.  We also discuss implications of the presence of free and dimensional damping parameters in the fCCZ4 formalism.
 
The paper is organized as follows. Section~\ref{section:form} describes the fCCZ4
evolution equations.  In Section~\ref{fCCCZ4-spherical} we write the fCCZ4 equations in
spherical coordinates under the assumption of spherical symmetry. 
Section~\ref{section:implementation} describes the numerical implementation and 
Section~\ref{section:results} shows results from a number of numerical experiments, namely 
a pure gauge wave, the evolution of a single black hole, the evolution of a spherical
relativistic star in equilibrium, the so-called migration test, and the gravitational collapse of 
a spherical relativistic star leading to the formation of a black hole.  We summarize and discuss
the respective advantages and disadvantages of fCCZ4 and BSSN
in Section~\ref{section:summary}.  Throughout this article we will 
use gravitational units $c=G=1$. Greek indices denote spacetime indices (0 to 3), while Latin indices denote space indices only (1 to 3).

\section{The fully Covariant and Conformal Z4 formulation}
\label{section:form}

The Z4 constraint damped system~\cite{Bona,Gundlach} in its 4-dimensional covariant form replaces the
Einstein equations by 
\begin{eqnarray} \label{einstein1}
{}^{(4)}R_{\mu \nu}+\nabla_{\mu} {}^{(4)} Z_{\nu}+\nabla_{\nu} {}^{(4)} Z_{\mu} -
\kappa_1[n_{\mu}{}^{(4)}Z_{\nu}+n_{\nu}{}^{(4)}Z_{\mu} \nonumber\\
-(1+\kappa_2)g_{\mu \nu}n_{\sigma}{}^{(4)}Z^{\sigma}]= 8\pi
\left(T_{\mu\nu}-\frac{1}{2}g_{\mu \nu}T\right),\quad
\end{eqnarray}
where ${}^{(4)}R_{\mu \nu}$ is the Ricci tensor of the 4-dimensional spacetime ${\cal M}$
with metric $g_{\mu \nu}$, $\nabla_{\mu}$ the covariant derivative
associated with metric $g_{\mu \nu}$, $T_{\mu \nu}$ the stress-energy tensor and
$T\equiv g_{\mu \nu} T^{\mu \nu}$ its trace.   The above equation reduces to Einstein's equations when the additional 4-vector ${}^{(4)}Z_{\mu}$ vanishes.  The two arbitrary constants $\kappa_1$ and $\kappa_2$ serve as constraint damping coefficients.  While $\kappa_2$ is dimensionless, $\kappa_1$ has units of inverse length.

In the 3+1 decomposition we assume that the spacetime ${\cal M}$ can be foliated by a family of
spatial slices $\Sigma$ that coincide with level surfaces of a
coordinate time $t$.   We denote the future-pointing unit normal on
$\Sigma$ with $n^\mu$ and write the line element as
\begin{eqnarray} \label{metric}
ds^2 & = & - \alpha^2 dt^2 + \gamma_{ij} (dx^i + \beta^i dt)(dx^j + \beta^j dt),
\end{eqnarray}
where $\alpha$ is the lapse function, $\beta^i$ the shift vector, and $\gamma_{ij}$ the spatial metric induced on $\Sigma$. In terms of the lapse and shift, the normal vector $n^\mu$ can be expressed as
\begin{equation} \label{normal}
n_\mu = (-\alpha,0,0,0)~~~\mbox{or}~~~n^\mu = (1/\alpha, - \beta^i/\alpha).
\end{equation}

As in the BSSN formulation we adopt a conformal decomposition of the spatial metric
\begin{equation} \label{conformal_decomposition}
\gamma_{ij} = e^{4 \phi} \bar \gamma_{ij},
\end{equation}
where $e^{4 \phi}$ is the conformal factor and $\bar \gamma_{ij}$ the conformally related metric.  We will refer to the connection coefficients associated with $\bar \gamma_{ij}$ as $\bar \Gamma^i_{jk}$.  Instead of determining the conformal factor by fixing the determinant of the conformal metric, $\bar \gamma$ to unity, as is suitable for Cartesian coordinates, we adopt
 \begin{equation}\label{eq3}
e^{4\phi}=(\gamma/\bar \gamma)^{1/3},
\end{equation}
where $\gamma$ is the determinant of $\gamma_{ij}$.   In order to determine the conformal factor we then impose Brown's ``Lagrangian" condition
\begin{equation} \label{Lagrange}
 \partial_{t}\bar\gamma=0.
\end{equation}

We denote the conformally rescaled extrinsic curvature as
\begin{equation}\label{eq4}
\bar A_{ij}=e^{-4\phi}\,\biggl(K_{ij}-\frac{1}{3}\gamma_{ij}K\biggl),
\end{equation}
where $K_{ij}$ is the physical extrinsic curvature and
$K=\gamma^{ij}K_{ij}$ its trace. 

We next introduce a reference metric $\hat
\gamma_{ij}$ with corresponding reference connection
$\hat\Gamma^{i}_{jk}$.   We then define the difference between the connections associated with the conformally related and the reference metric as
\begin{equation}\label{eq5}
\Delta\Gamma^{i}_{jk} \equiv \bar\Gamma^{i}_{jk}-\hat\Gamma^{i}_{jk},
\end{equation}
and note that, unlike the individual connections, these objects transform as a tensor field. 

In the Z4 system, the Hamiltonian and momentum constraints
result in equations for the four-vector ${}^{(4)}Z_{\mu}$.  In a 3+1 decomposition, these equations can be written as evolution equations for the projection of the ${}^{(4)}Z_\mu$ along the normal $n^\mu$, which, following convention, we define as
\begin{equation}
\Theta \equiv -n_{\mu}{}^{(4)}Z^{\mu}=\alpha {}^{(4)}Z^{0},
\end{equation}
and the spatial projection of ${}^{(4)}Z_{\mu}$,
\begin{equation}
Z_{i}\equiv \gamma_{i}^{~\mu}{}^{(4)}Z_{\mu}.
\end{equation}
Here $Z_i$ now denotes a spatial vector whose index can be raised with the (inverse) spatial metric, $Z^i = \gamma^{ij} Z_j$.

Defining
\begin{equation}\label{eq7}
\partial_{\bot}\equiv\partial_{t}-\mathcal{L}_{\beta}
\end{equation}
where $\mathcal{L}_{\beta}$ denotes the Lie derivative along the shift vector $\beta^{i}$, the fully covariant and conformal Z4 system in a reference-metric approach (fCCZ4) is then given by the following set of evolution equations:
\begin{eqnarray}
\partial_{\bot}\bar\gamma_{ij}&=&-\frac{2}{3}\bar\gamma_{ij}\mathcal{\bar D}_{k}\beta^{k}-2\alpha\bar A_{ij},\label{eq8}\\
\partial_{\bot}\bar A_{ij}&=&-\frac{2}{3}\bar A_{ij}\mathcal{\bar D}_{k}\beta^{k}-2\alpha\bar A_{ik}\bar A^{k}_{j}+\alpha\bar A_{ij}(K-2\Theta)\nonumber\\
&&+e^{-4\phi}\bigl[-2\alpha\mathcal{\bar D}_{i}\mathcal{\bar D}_{j}\phi+4\alpha\mathcal{\bar D}_{i}\phi\mathcal{\bar D}_{j}\phi\nonumber\\
&&+4\mathcal{\bar D}_{(i}\alpha\mathcal{\bar D}_{j)}\phi-\mathcal{\bar D}_{i}\mathcal{\bar D}_{j}\alpha\nonumber\\
&&+\alpha(\bar R_{ij}+\mathcal{D}_{i}Z_{j}+\mathcal{D}_{j}Z_{i}-8\pi S_{ij})\bigl]^{\text{TF}},\label{eq9}\\
\partial_{\bot}\phi&=&\frac{1}{6}\mathcal{\bar D}_{i}\beta^{i}-\frac{1}{6}\alpha K,\label{eq10}\\
\partial_{\bot}K&=&e^{-4\phi}\bigl[\alpha\bigl(\bar R-8\mathcal{\bar D}^{i}\phi\mathcal{\bar D}_{i}\phi-8\mathcal{\bar D}^{2}\phi\bigl)\nonumber\\
&&-\bigl(2\mathcal{\bar D}^{i}\alpha\mathcal{\bar D}_{i}\phi+\mathcal{\bar D}^{2}\alpha\bigl)\bigl]+\alpha(K^{2}-2\Theta K)\nonumber\\
&&+2\alpha\mathcal{D}_{i}Z^{i}-3\alpha\kappa_{1}(1+\kappa_{2})\Theta\nonumber\\
&&+4\pi\alpha(S-3E),\label{eq11}\\
\partial_{\bot}\Theta&=&\frac{1}{2}\alpha\bigl[e^{-4\phi}\bigl(\bar R-8\mathcal{\bar D}^{i}\phi\mathcal{\bar D}_{i}\phi-8\mathcal{\bar D}^{2}\phi\bigl)\nonumber\\
&&-\bar A^{ij}\bar A_{ij}+\frac{2}{3}K^{2}-2\Theta K+2\mathcal{D}_{i}Z^{i}\bigl]\nonumber\\
&&-Z^{i}\partial_{i}\alpha-\alpha\kappa_{1}(2+\kappa_{2})\Theta-8\pi\alpha E,\label{eq12}\\
\partial_{\bot}\tilde\Lambda^{i}&=&\bar\gamma^{jk}\mathcal{\hat D}_{j}\mathcal{\hat D}_{k}\beta^{i}+\frac{2}{3}\Delta\Gamma^{i}\mathcal{\bar D}_{j}\beta^{j}+\frac{1}{3}\mathcal{\bar D}^{i}\mathcal{\bar D}_{j}\beta^{j}\nonumber\\
&&-2\bar A^{jk}\bigl(\delta^{i}_{\,\,j}\partial_{k}\alpha-6\alpha\delta^{i}_{\,\,j}\partial_{k}\phi-\alpha\Delta\Gamma^{i}_{jk}\bigl)\nonumber\\
&&-
\frac{4}{3}\alpha\bar\gamma^{ij}\partial_{j}K+2\bar\gamma^{ki}\bigl(\alpha\partial_{k}\Theta-\Theta\partial_{k}\alpha-\frac{2}{3}\alpha K Z_{k}\bigl)\nonumber\\
&&-2\alpha\kappa_{1}\bar\gamma^{ij}Z_{j}-16\pi\alpha\bar\gamma^{ij}S_{j}.\label{eq13}
\end{eqnarray}
Here the superscript TF denotes the trace-free part of a tensor, $\kappa_1$ and $\kappa_2$ are the damping coefficients introduced by~\cite{Gundlach}, and $\mathcal{\hat D}_{i}$, $\mathcal{D}_{i}$ 
and $\mathcal{\bar D}_{i}$ denote the covariant derivatives built from the connection associated 
with the reference metric $\hat\gamma_{ij}$, the physical metric $\gamma_{ij}$ and the conformal 
metric $\bar\gamma_{ij}$, respectively.  We have also defined 
\begin{equation} \label{def_Lambda}
\tilde\Lambda^{i}\equiv\bar\Lambda^{i}+2\bar\gamma^{ij}Z_{j}, 
\end{equation}
where
\begin{equation}\label{eq6}
\bar\Lambda^{i}\equiv \Delta \Gamma^i = \bar\gamma^{jk}\Delta\Gamma^{i}_{jk}.
\end{equation}
The vector $\tilde\Lambda^{i}$ plays the role of the ``conformal connection functions'' in the original
CCZ4 system; its evolution equation (\ref{eq13}) involves the evolution equation for the variables $Z_i$.  

The matter sources $E$, $S_{i}$, $S_{ij}$ and $S$ denote the density, momentum density, stress, and the trace of the stress as observed by a normal observer, respectively:
\begin{eqnarray}
E&\equiv& n_{\mu}n_{\nu}T^{\mu\nu},\label{eq17}\\
S_{i}&\equiv&-\gamma_{i\mu}n_{\nu}T^{\mu\nu},\label{eq18}\\
S_{ij}&\equiv&\gamma_{i\mu}\gamma_{j\nu}T^{\mu\nu},\label{eq19}\\
S&\equiv&\gamma^{ij}S_{ij}.\label{eq20}
\end{eqnarray}

In Eq.~(\ref{eq9}), we compute the Ricci tensor $\bar R_{ij}$ associated with $\bar\gamma_{ij}$ from 
\begin{equation}\label{eq24}
\begin{split}
\bar R_{ij}= &-\frac{1}{2}\bar\gamma^{kl}\mathcal{\hat D}_{k}\mathcal{\hat D}_{l}\bar\gamma_{ij}+\bar\gamma_{(i}\mathcal{\hat D}_{j)}\Delta\Gamma^{k}+\Delta\Gamma^{k}\Delta\Gamma_{(ij)k}\\
&+\bar\gamma^{kl}\bigl(2\Delta\Gamma^{m}_{k(i)}\Delta\Gamma_{j)ml}+\Delta\Gamma^{m}_{ik}\Delta\Gamma_{mjl}\bigl).
\end{split}
\end{equation}
Here we compute the $\Delta \Gamma^i$ from their definition (\ref{eq6}).  Given $\Delta \Gamma^i$, and values for $\tilde \Lambda^i$, the vectors $Z_i$, which are not evolved independently, can be determined from (\ref{def_Lambda}).

Unless stated otherwise we fix the gauge freedom by imposing the so called ``non-advective 1+log"
condition for the lapse~\cite{Bona97a} 
\begin{equation}
\partial_{t}\alpha = -2\alpha(K-2\Theta),\label{eq14}
\end{equation}
and a variation of the
"Gamma-driver" condition for the shift vector~\cite{Alcubierre02a}
\begin{eqnarray}
\partial_{t}\beta&=&B^{i},\label{eq15}\\
\partial_{t}B^{i}&=&\frac{3}{4}\partial_{t}\tilde\Lambda^{i}\label{eq16}.
\end{eqnarray}

Finally, when $\Theta = Z_{i} = 0$, the 
evolution equations (\ref{eq8})-(\ref{eq13}) imply that the Hamiltonian and momentum 
constraints hold in the form
\begin{eqnarray}
\mathcal{H}&\equiv&\frac{2}{3}K^{2}-\bar A_{ij}\bar A^{ij}+e^{-4\phi}\bigl(\bar R-8\mathcal{\bar D}^{i}\phi\mathcal{\bar D}_{i}\phi-8\mathcal{\bar D}^{2}\phi\bigl)\nonumber\\
&&-16\pi E=0,\label{eq25}\\
\mathcal{M}^{i}&\equiv&e^{-4\phi}\bigl(\frac{1}{\displaystyle\sqrt{\bar\gamma}}\mathcal{\hat D}_{j}(\sqrt{\bar\gamma}\bar A^{ij})+6\bar A^{ij}\partial_{j}\phi-\frac{2}{3}\bar\gamma^{ij}\partial_{j}K\nonumber\\
&&+\bar A^{jk}\Delta\Gamma^{i}_{jk}\bigl)-8\pi S^{i}=0,\label{eq26}
\end{eqnarray}
where ${\bar R}$ is the trace of $\bar R_{ij}$.

In Cartesian coordinates, when $\bar\gamma=1$ and $\hat\Gamma^{i}_{jk}=0$, the above equations reduce to the CCZ4 equations of \cite{Alic1}, except that we have set their coefficients $\kappa_3$ to unity.

\section{Spherical symmetry}
\subsection{The $\text{f}$CCZ4 equations}
\label{fCCCZ4-spherical}

Under the assumption of spherical symmetry, the space line element can
be written in spherical coordinates ($r,\theta,\varphi$) as 
\begin{equation}\label{eq27}
dl^{2}=e^{4\phi}[a(r,t)dr^{2}+r^{2}b(r,t)d\Omega^{2}],
\end{equation}
where $d\Omega^{2}=d\theta^{2}+\sin^{2}\theta d\varphi^{2}$ is the solid angle element, and $a(r,t)$ and
$b(r,t)$ are the metric functions.  Since the evolution equations for the conformally related metric and the conformal factor, eqs.~(\ref{eq8}) and (\ref{eq10}), take the exact from as their counterparts in the BSSN formulation, their spherically symmetric versions also remain unchanged
\begin{eqnarray}
\partial_{t}X&=&\beta^{r}\partial_{r}X-\frac{1}{3}X\sigma\mathcal{\bar D}_{m}\beta^{m}+\frac{1}{3}X\alpha K,\label{eq28}\\
\partial_{t}a&=&\beta^{r}\partial_{r}a+2a\partial_{r}\beta^{r}-\frac{2}{3}\sigma a\mathcal{\bar D}_{m}\beta^{m}-2\alpha aA_{a},\label{eq29}\\
\partial_{t}b&=&\beta^{r}\partial_{r}b+2b\frac{\beta^{r}}{r}-\frac{2}{3}\sigma b\mathcal{\bar D}_{m}\beta^{m}-2\alpha bA_{b},\label{eq30}
\end{eqnarray}
(see \cite{Alcubierre} for the BSSN system in spherical symmetry.)  Here $X\equiv e^{-2\phi}$ and $\sigma=1$ to impose the Lagrangian condition (\ref{Lagrange}) on the time evolution of the determinant of the conformal metric.   The covariant derivative of 
the shift vector can be written as
\begin{equation}\label{eq31}
\mathcal{\bar D}_{m}\beta^{m}=\partial_{r}\beta^{r}+\beta^{r}\biggl(\frac{\partial_{r}(ab^{2})}{2ab^{2}}+\frac{2}{r}\biggl),
\end{equation}
and we have defined
\begin{equation}\label{eq32}
A_{a}\equiv \bar A^{r}_{r},\quad A_{b}\equiv \bar A^{\theta}_{\theta}.
\end{equation}
Note that the quantity $X$ is evolved in Eq.~(\ref{eq28}) instead of the conformal factor $\phi$ itself.

The evolution equation for the trace of the extrinsic curvature $K$ is
\begin{eqnarray}\label{eq33}
\partial_{t}K&=&-\mathcal{D}^{2}\alpha+\alpha\bigl(R+2\mathcal{D}_{m}Z^{m}+K^{2}-2\Theta K\bigl)+\beta^{r}\partial_{r} K\nonumber \\
&-&3\alpha\kappa_{1}(1+\kappa_{2})\Theta+4\pi\alpha\bigl(S_{a}+2S_{b}-3 E \bigl),
\end{eqnarray}
while for $\Theta$ we have
\begin{eqnarray}\label{eq34}
\partial_{t}\Theta&=&\frac{1}{2}\alpha\bigl(R+2\mathcal{D}_{m}Z^{m}-(A_{a}^{2}+2A_{b}^{2})+\frac{2}{3}K^{2}-2\Theta K\bigl)\nonumber \\
&+&\beta^{r}\partial_{r}\Theta-Z^{r}\partial_{r}\alpha-\alpha\kappa_{1}(2+\kappa_{2})\Theta-8\pi\alpha E.
\end{eqnarray}
Here we defined $S_{a}\equiv S^{r}_{r}$ and $S_{b}\equiv S^{\theta}_{\theta}$. The divergence of the $Z_{i}$ vector with respect to the physical metric is 
\begin{equation}\label{eq35}
\mathcal{D}_{m}Z^{m}=\partial_{r}Z^{r}+Z^{r}\biggl(\frac{\partial_{r}(ab^{2})}{2ab^{2}}+\frac{2}{r}+6\partial_{r}\phi\biggl).
\end{equation}

In spherical symmetry, the evolution equation (\ref{eq9}) for the independent component of the traceless part of the conformal extrinsic curvature, $A_{a}$,  reduces to
\begin{eqnarray}\label{eq36}
\partial_{t}A_{a}&=&\beta^{r}\partial_{r}A_{a}-\bigl(\mathcal{D}^{r}\mathcal{D}_{r}\alpha-\frac{1}{3}\mathcal{D}^{2}\alpha\bigl)+\alpha\bigl(R^{r}_{r}-\frac{1}{3}R\bigl) \nonumber \\
&+&\alpha\bigl(2\mathcal{D}_{r}Z^{r}-\frac{2}{3}\mathcal{D}_{m}Z^{m}\bigl) \nonumber \\
&+&\alpha A_{a}(K-2\Theta)-16\pi\alpha(S_{a}-S_{b}),
\end{eqnarray}
where $R^{r}_{r}$ is the mixed radial component of the physical, spatial Ricci tensor. The covariant derivative of the $Z_{r}$ is
\begin{gather}
\mathcal{D}_{r}Z^{r}=\bigl[\partial_{r}Z^{r}+Z^{r}\bigl(\frac{\partial_{r}a}{2a}+2\partial_{r}\phi\bigl)\bigl]\label{38}.
\end{gather}

From the definition (\ref{def_Lambda}) we have
\begin{equation}\label{40}
\tilde\Lambda^{r}\equiv\bar\Lambda^{r}+\frac{2}{a}Z_{r},
\end{equation}
where
\begin{equation}\label{41}
\bar\Lambda^{r}=\frac{1}{a}\biggl[\frac{\partial_{r}a}{2a}-\frac{\partial_{r}b}{b}-\frac{2}{r}\biggl(1-\frac{a}{b}\biggl)\biggl].
\end{equation}
The evolution equation for $\tilde\Delta^{r}$ in spherical symmetry can then be derived from Eq.~(\ref{eq13}),
\begin{equation}\label{eq39}
\begin{split}
\partial_{t}\tilde\Lambda^{r}&=\beta^{r}\partial_{r}\tilde\Lambda^{r}-\bar\Lambda^{r}\partial_{r}\beta^{r}+\frac{1}{a}\partial^{2}_{r}\beta^{r}+\frac{2}{b}\partial_{r}\biggl(\frac{\beta^{r}}{r}\biggl)\\
&+\frac{\sigma}{3}\biggl(\frac{1}{a}\partial_{r}(\mathcal{\bar D}_{m}\beta^{m})+2\bar\Lambda^{r}\mathcal{\bar D}_{m}\beta^{m}\biggl)\\
&-\frac{2}{a}(A_{a}\partial_{r}\alpha+\alpha\partial A_{a})\\
&+2\alpha\biggl(A_{a}\bar\Lambda^{r}-\frac{2}{rb}(A_{a}-A_{b})\biggl)\\
&+\frac{2\alpha}{a}\biggl[\partial_{r}A_{a}\frac{2}{3}\partial_{r}K+6A_{a}\partial_{r}\phi\\
&+(A_{a}-A_{b})\biggl(\frac{2}{r}+\frac{\partial_{r}b}{b}\biggl)-8\pi S_{r}\biggl]\\
&+\frac{2}{a}\biggl(\alpha\partial_{r}\Theta-\Theta\partial_{r}\alpha-\frac{2}{3}\alpha KZ_{r}\biggl)\\
&+\frac{2}{a}\biggl(\frac{2}{3}Z_{r}\mathcal{\bar D}_{m}\beta^{m}-Z_{r}\partial_{r}\beta^{r}\biggl)-\frac{2}{a}\kappa_{1}Z_{r}.
\end{split}
\end{equation}

The Hamiltonian and momentum constrains are given by the following two equations that we compute 
to monitor the accuracy of the numerical evolutions:
\begin{eqnarray}
\mathcal{H}&\equiv& R-(A^{2}_{a}+2A_{b}^{2})+\frac{2}{3}K^{2}-16\pi E=0,\label{eq42}\\
\mathcal{M}^{r}&\equiv&\partial_{r}A_{a}-\frac{2}{3}\partial_{r}K+6A_{a}\partial_{r}\phi\nonumber\\
&+&(A_{a}-A_{b})\biggl(\frac{2}{r}+\frac{\partial_{r}b}{b}\biggl)-8\pi S_{r}=0.\label{eq43}
\end{eqnarray}

The gauge condition for the lapse and the shift are the same as in
Eqs.~(\ref{eq14}-\ref{eq16}), but taking only the radial component for
the shift and the vector $B^{i}$, and replacing $\tilde\Lambda^{i}$ by $\tilde\Lambda^{r}$ as in the evolution equation (\ref{eq39}).

\subsection{Hydrodynamics}

The general relativistic hydrodynamics equations, expressed through the 
conservation equation for the stress-energy tensor $T^{\mu\nu}$ and the 
continuity equation, are
\begin{equation}
\label{hydro eqs}
	\nabla_\mu T^{\mu\nu} = 0\;,\;\;\;\;\;\;
\nabla_\mu \left(\rho u^{\mu}\right) = 0,
\end{equation}
where $\rho$ is the rest-mass density and $u^{\mu}$ the 4-velocity of the fluid.
Following \cite{Banyuls97}, we write the equations of general relativistic hydrodynamics
in a conservative form in spherical symmetry.  We define the fluid 3-velocity as seen by a normal observer as
\begin{equation}
	v^{r}\equiv \frac{u^{r}}{\alpha u^{t}}+\frac{\beta^{r}}{\alpha},
\end{equation}
and the Lorentz factor between the fluid and the normal observer as
\begin{equation}
\label{def2}
	W\equiv \alpha u^{t}.
\end{equation}
 We also define the fluid density, momentum density and energy density, all as observed by a normal observer, as
\begin{align}
	D & = \rho W,\\
	S_r & = \rho h W^2v_r,\\
	\tau & = \rho h W^2 - P - D,
\end{align}
where $h$ is the specific enthalpy and $P$ the pressure.  We then assemble these variables into a vector
${\bf{U}}$ of conserved fluid variables 
\begin{equation}
	{\bf{U}}=\sqrt{\gamma}(D,S_{r},\tau).
\end{equation}
Defining corresponding fluxes, ${\bf{F}}^{r}$, as
\begin{align}
{\bf{F}}^{r} & =\sqrt{-g}\left[D(v^{r}
  -\beta^{r}/\alpha), \right. \nonumber \\
&  \left. S_{r}(v^{r}-\beta^{r}/\alpha)+P, \right. \nonumber \\
& \left. \tau  (v^{r}-\beta^{r}/\alpha)+P v^r\right],
\end{align}
we can cast the equations of hydrodynamics  (\ref{hydro eqs}) 
in conservative form
\begin{equation}
\label{cylcon}
	\partial_{t}{\bf{U}}+\partial_{r}{\bf{F}}^{r}={\bf{S}}.
\end{equation}
Here ${\bf{S}}$ is a vector of source terms given by 
\begin{align}
	{\bf{S}} &= \sqrt{-g}\left[0, 
T^{00}\left(\frac{1}{2}(\beta^{r})^{2}\partial_{r}\gamma_{rr}
- \alpha\partial_{r}\alpha \right) \right. \nonumber\\
 & \left. + T^{0r}\beta^r\partial_{r}\gamma_{rr}+T^{0}_{r}\partial_{r}\beta^{r}
  \frac{1}{2}T^{rr}\partial_{r}\gamma_{rr}, \right. \nonumber\\ 
 & (T^{00}\beta^{r}+T^{0r})(\beta^{r}K_{rr} - \partial_{r}\alpha)+T^{rr}K_{rr}
  \bigg].
\end{align}
To close the system of equations, we choose a Gamma-law equation of state
(EOS) 
\begin{equation}
\label{EOS1}
	P=\left(\Gamma -1\right)\rho\epsilon ,
\end{equation}
where $\Gamma$ is the adiabatic index and $\epsilon$ is the specific internal energy.

\section{Numerical Implementation}
\label{section:implementation}
\subsection{PIRK method}

We have implemented the fCCZ4 system under the assumption of spherical symmetry in the 1D-code 
described in Montero and Cordero-Carri\'on \cite{Pedro1}. This code solves the Einstein
equations coupled to the general relativistic hydrodynamics equations. The Einstein
equations are solved using either the BSSN or the fCCZ4
formalisms. We employ a second-order PIRK method to integrate the
evolution equations in time. Writing a system of PDEs as follows  
%
%
\begin{System}
u_t = \mathcal{L}_1 (u, v), \\
v_t = \mathcal{L}_2 (u) + \mathcal{L}_3 (u, v),
\label{e:system}
\end{System}
where $\mathcal{L}_1$, $\mathcal{L}_2$ and $\mathcal{L}_3$ represent general non-linear 
differential operators, the second-order PIRK method takes the following form:
\begin{System}
	u^{(1)} = u^n + \Delta t \, L_1 (u^n, v^n),  \\
	v^{(1)} = v^n + \Delta t \left[\frac{1}{2} L_2(u^n) +
          \frac{1}{2} L_2(u^{(1)}) + L_3(u^n, v^n) \right],
\end{System}
\begin{System}
	u^{n+1}  = \frac{1}{2} \left[ u^n + u^{(1)} 
+ \Delta t \, L_1 (u^{(1)}, v^{(1)}) \right], \\
	v^{n+1}  =   v^n + \frac{\Delta t}{2} \left[ 
L_2(u^n) + L_2(u^{n+1}) \right.  \\ 
\left. \hspace{2.5cm} +  L_3(u^n, v^n) + L_3 (u^{(1)}, v^{(1)}) \right],
\end{System}
where we denote by $L_1$, $L_2$ and $L_3$ the corresponding discrete
operators. In particular, we note that $L_1$ and $L_3$ will be treated in an explicit way, 
whereas the $L_2$ operator will contain the singular terms  appearing in the sources of 
the equations and, therefore, will be treated partially implicitly.

In the first stage, $u$ is evolved explicitly; the updated value $u^{(1)}$ is 
used in the evaluation of the $L_2$ operator for the computation of $v^{(1)}$.
Once all the values of the first stage are obtained, $u$ is evolved explicitly (using the values of the variables of the
previous time-step and previous stage), and the updated value $u^{n+1}$ is used
in the evaluation of the $L_2$ operator for the computation of $v^{n+1}$.

The precise evolution algorithm we use in the code is as follows:
\begin{itemize}
\item
Firstly, the hydrodynamic conserved quantities, the 
conformal metric components $a$ and $b$, the conformal factor $\phi$ or the 
quantity $X$, the lapse function $\alpha$, 
and the radial component of the shift vector $\beta^r$, are evolved explicitly (as 
$u$ is evolved in the previous PIRK scheme). 
\item
Secondly, the traceless part of 
the extrinsic curvature, $A_a$, the trace of the extrinsic curvature, $K$, and the projection of the four-vector $Z^{\mu}$ along the normal direction, $\Theta$,
are evolved partially implicitly, using updated values of $\alpha$, $a$, $b$ and $X$.
\item
Next, the quantity $\tilde{\Lambda}^{r}$ is evolved partially
implicitly using the updated values of $\alpha$, $a$, $b$, $\beta^r$, $X$, 
$A_a$, $K$ and $\Theta$.
\item
Finally, $B^r$ is evolved partially implicitly using the 
updated values of $\hat{\Lambda}^{r}$. 
\end{itemize}

We note that the matter source terms are always included 
in the explicitly treated parts. In Appendix~\ref{appendix}, we give the exact 
form of the source terms included in each operator.

\subsection{Numerics}
The spatial domain for our computations is defined as $0\leq r \leq L$, where $L$ refers to 
the location of the outer boundary. We use a cell-centered grid to avoid 
the origin from coinciding with a grid point.  At the origin we impose boundary conditions derived from
the assumption of spherical symmetry, while at the outer boundary we impose Sommerfeld
boundary conditions for the spacetime variables~\cite{Alcubierre02a,Pedro1}. 

We compute derivatives in the spacetime evolution using a 
fourth-order centered finite difference approximation on a uniform grid except 
for the advection terms (i.e.~terms of the form $\beta^{r}\partial_{r}u$), for 
which we adopt a fourth-order upwind scheme.  We also use fourth-order Kreiss-Oliger 
dissipation \cite{Kreiss} to avoid high frequency noise appearing near the 
outer boundary. 

For the equations of hydrodynamics we implement a high
resolution shock capturing scheme (HRSC) that consist of a second-order 
slope limiter reconstruction scheme (MC limiter) to obtain the left and right states of 
the primitive variables at each cell interface, and the HLLE approximate
Riemann solver~\citep{Harten83,Einfeldt88}.  We add a low density atmosphere to handle vacuum regions; more specifically
we treat the atmosphere as a perfect fluid with rest-mass density several orders of magnitude smaller than that of the bulk matter. Further details of our implementation can be found in \cite{Pedro1}.

\section{Numerical experiments}
\label{section:results}

We now describe several numerical experiments with the fCCZ4 formalism.  For each one we will describe the initial data for the gravitational field and hydrodynamics quantities in the corresponding Section; in addition we always impose $\Theta = 0$ and $Z_r=0$ at the initial time $t=0$. 

\subsection{Pure gauge dynamics}

We first consider the propagation of a pure gauge pulse.  Following \cite{Alcubierre,Pedro1} we choose as initial data 
\begin{eqnarray}
\phi&=&A_{a}=A_{b}=K=\tilde\Lambda^{r}=0,\label{eq44}\\
a&=&b=1,\label{eq45}\\
\alpha&=&1+\frac{\alpha_{0}r^{2}}{\lambda^2+r^{2}}\bigl[e^{-(r-r_{0})^{2}/\lambda^2}+e^{-(r+r_{0})^2/\lambda^2}\bigl],\label{eq48}
\end{eqnarray}
with $\alpha_{0}=0.01$ and $r_{0}=5 \lambda$. The quantity $\lambda$ is the length scale of the test.  In this test, we employ zero shift and harmonic slicing. The slicing condition is suitably written for the fCCZ4 formulation by introducing
the $\Theta$ variable in the evolution equation for the lapse function,
\begin{equation}\label{eq48b}
\partial_{t}\alpha=-\alpha^{2}(K-2\Theta).
\end{equation}

We choose a grid resolution of $\Delta r=0.1 \lambda$ (except for the convergence test described at the end of this Section) and a time step of
$\Delta t= C \Delta r$, where $C$ is the Courant factor.   
Among our first observations is that the fCCZ4 formalism requires a smaller Courant factor than the BSSN formalism, confirming similar findings by \cite{Alic2}; we found stable evolution for $C = 0.3$ for fCCZ4, and $C = 0.5$ for BSSN.  

\begin{figure}
\begin{minipage}{1\linewidth}
  \vspace{-0cm}\includegraphics[width=0.57\textwidth]{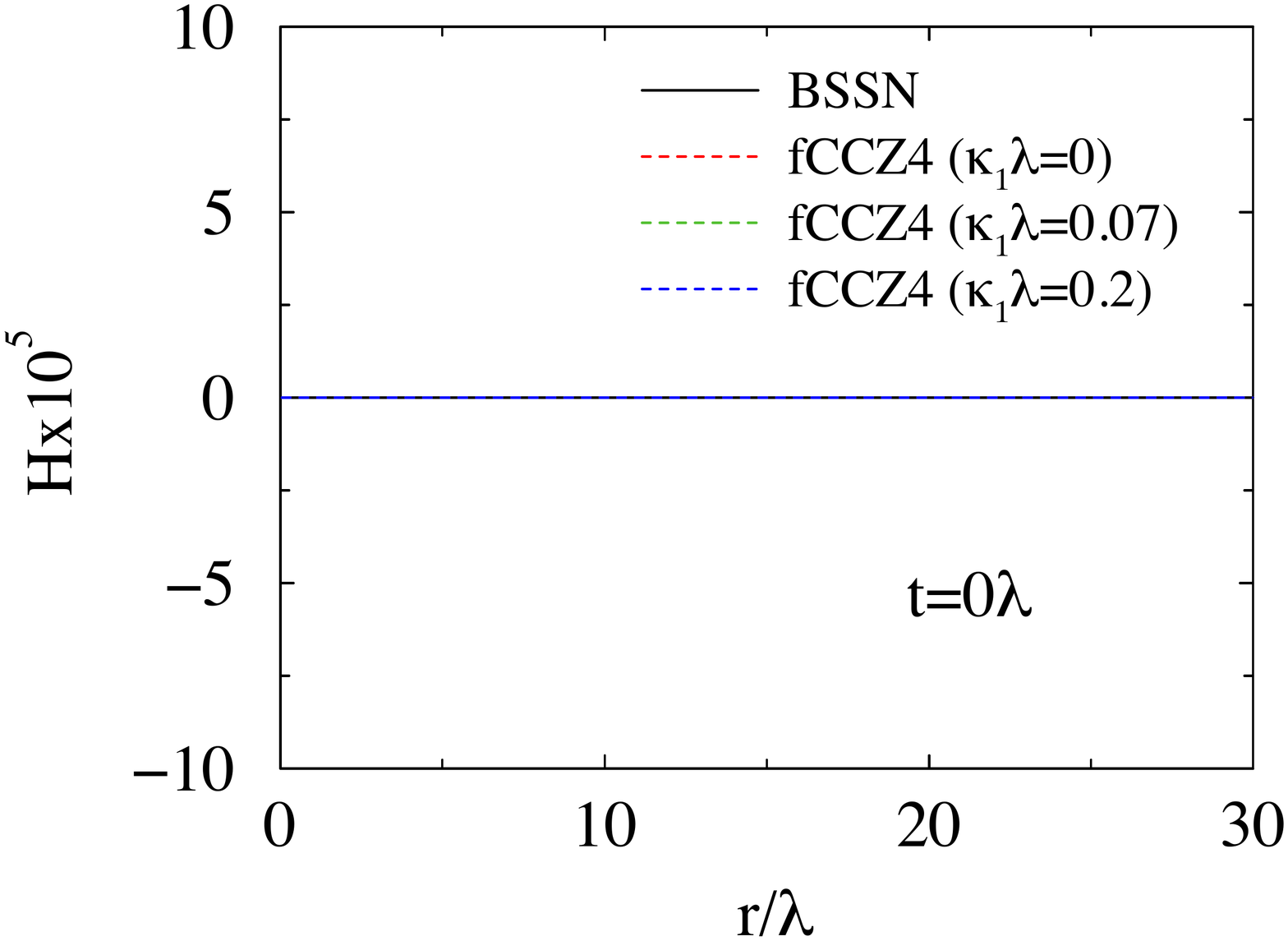}
  \hspace{-1.4cm}\includegraphics[width=0.57\textwidth]{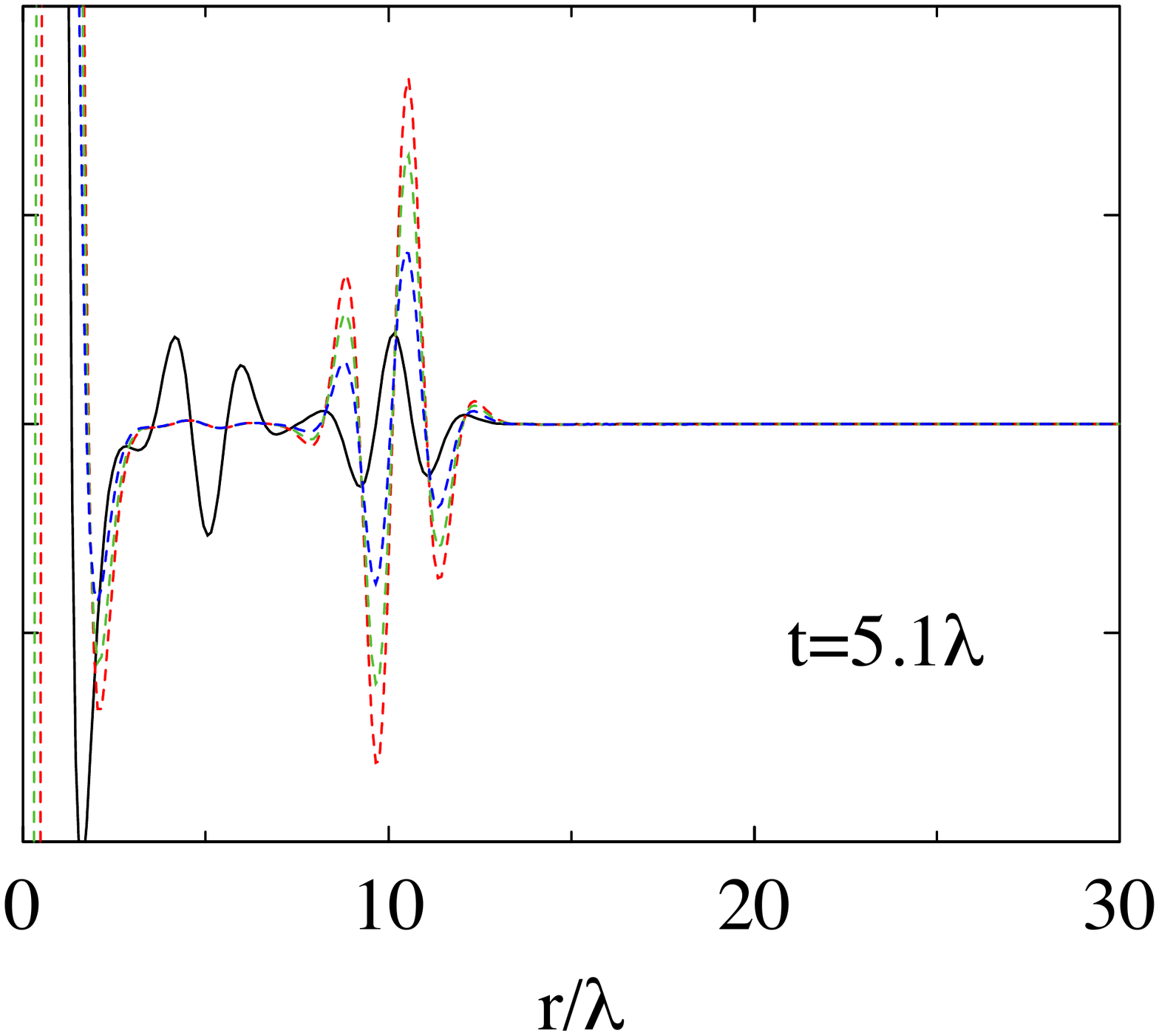}
\vspace{-0.6cm}
\\
\includegraphics[width=0.57\textwidth]{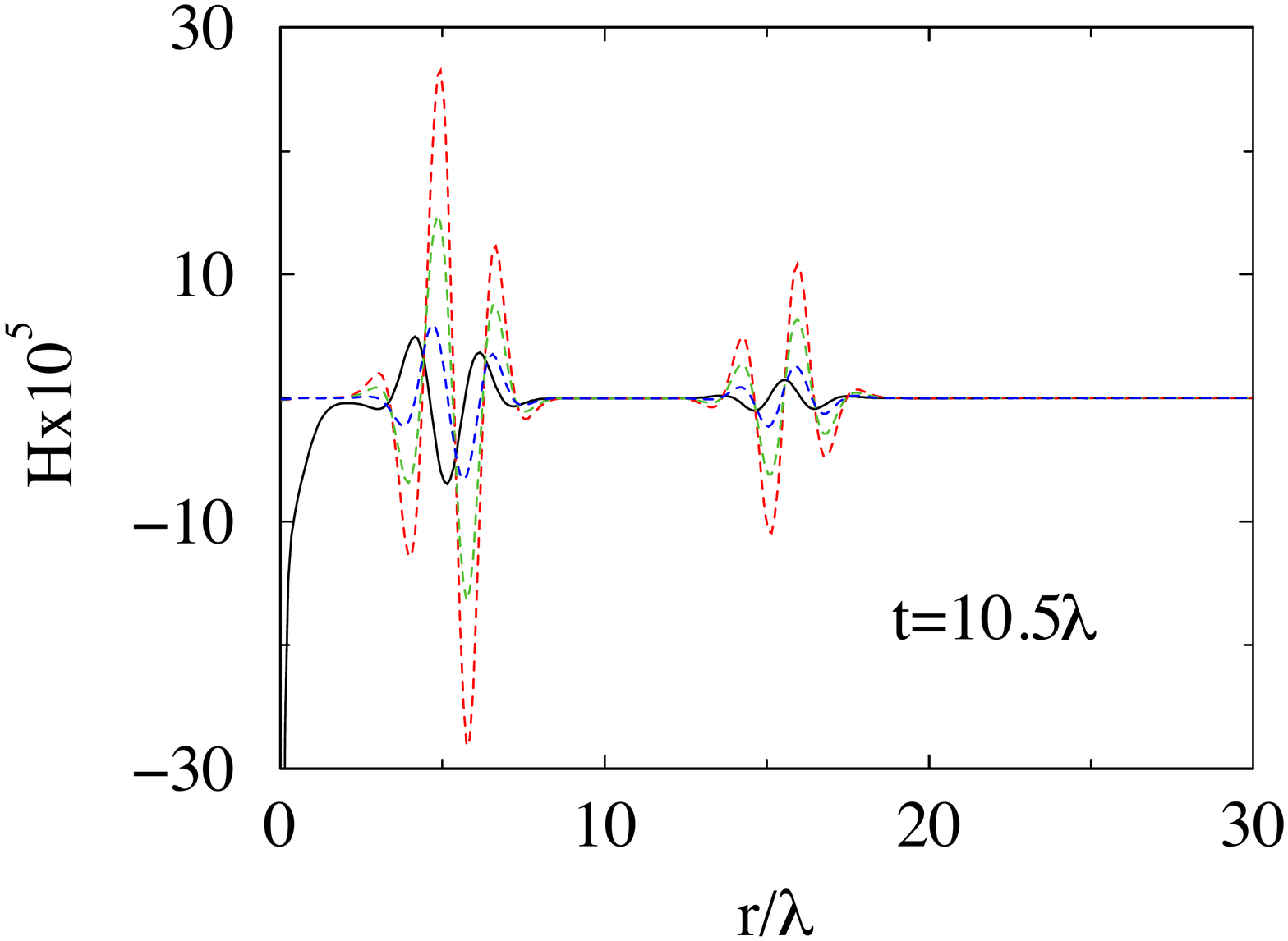}
\hspace{-1.4cm}\includegraphics[width=0.57\textwidth]{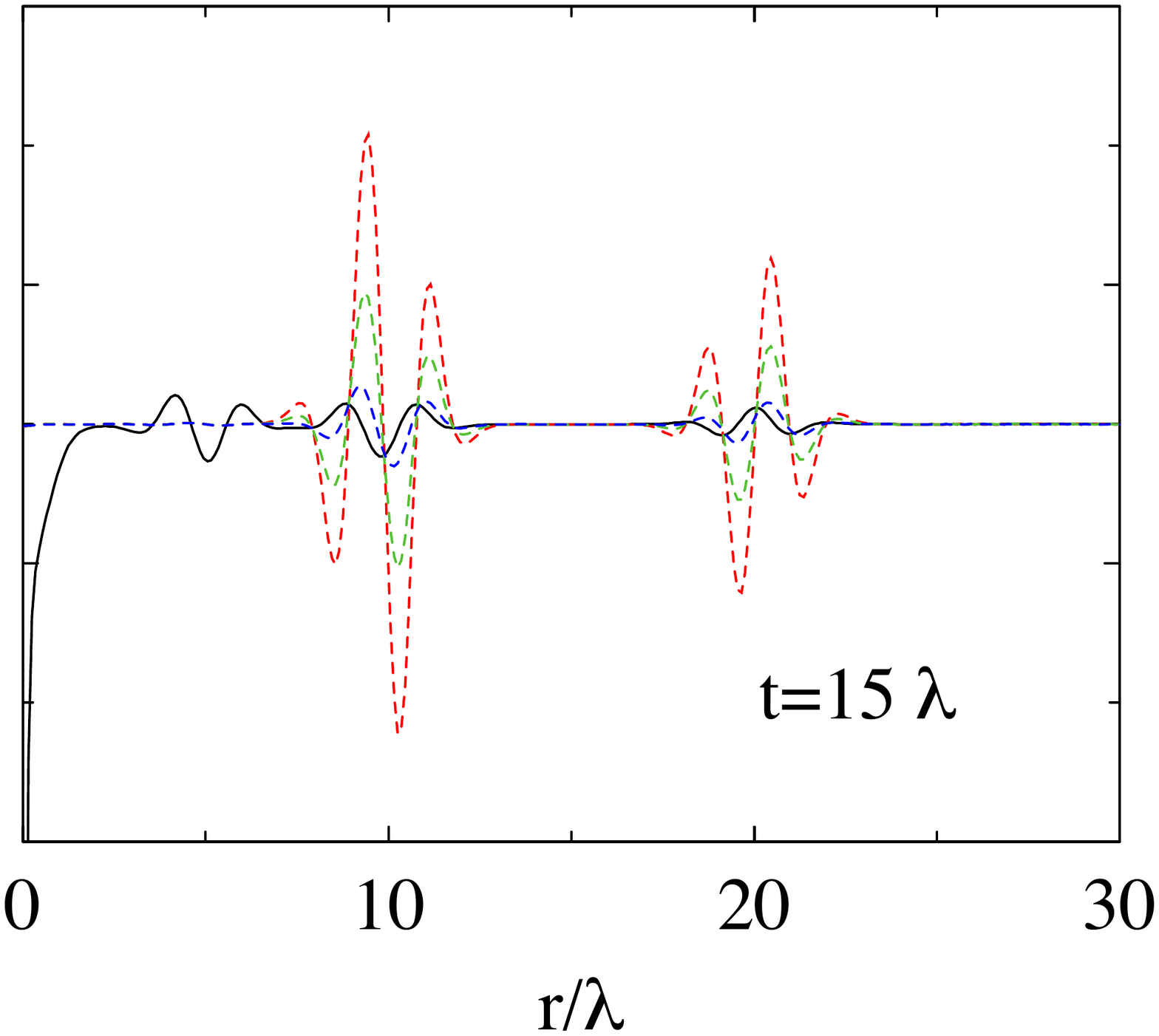}
\caption{ Hamiltonian constraint violation for a pure gauge pulse test as 
a function of radius at four different times for both BSSN (solid 
line) and fCCZ4 (dashed lines).}
\label{fig:graf1}
\end{minipage}
\end{figure}

In Fig.~\ref{fig:graf1} we show the Hamiltonian constraint at four
different times ($t/\lambda=0,5.1,10.5,15$) for evolutions performed with the
BSSN and the fCCZ4 formulations.   Following~\cite{Alic2} 
and~\cite{Sebastiano1} we also compare different choices for the damping parameters
$\kappa_{1} \lambda =\lbrace0,0.02,0.07,0.2\rbrace$ and $\kappa_{2}=\lbrace-0.5,0.5\rbrace$ for the fCCZ4 system. 
For the BSSN system, the violations of the Hamiltonian constraint settle down to approximately $10^{-3}$ close 
to the origin at $r=0$, and do not decrease with time after that (recall that we do not employ any 
regularization scheme at the origin).  As shown in Fig.~\ref{fig:graf1}, the behavior for the fCCZ4 system is different.  Here, 
the constraint violations propagate toward the outer boundary; close to the origin, the constraint violations end up being
approximately three order of magnitude smaller than for the BSSN system. We also note that the constraint violations decrease
with increasing values of the damping parameter $\kappa_{1}$.  However,
one should handle this parameter with precaution as we observed that
taking larger values for $\kappa_{1}$ (e.g. $\kappa_{1}/\lambda=5$) reduces
the propagation of the Hamiltonian constraint violation considerably and
leads to over-damping effects: a ``pulse"  remains near the origin.

\begin{figure}
\begin{minipage}{1\linewidth}
  \hspace{-0.5cm}\includegraphics[width=1.11\textwidth, height=0.3\textheight]{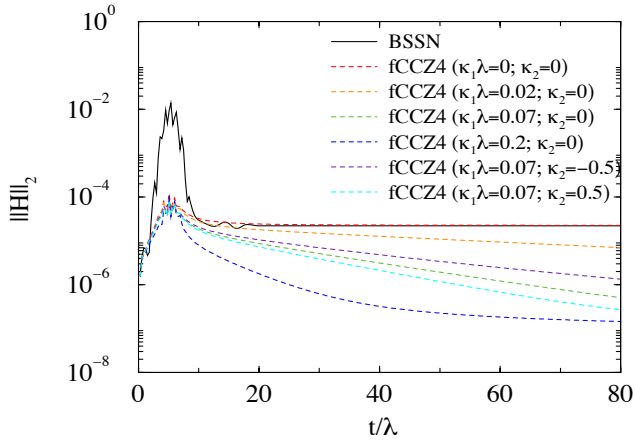}\vspace{-0.5cm}
\caption{L2-norm of the Hamiltonian constraint for a pure gauge pulse test for BSSN 
(solid line) and fCCZ4 (dashed lines) as a function of time and for different choices
of the damping parameters.}
\label{fig:graf2}
\end{minipage}
\end{figure}

In Fig.~\ref{fig:graf2} we plot the L2-norm, which is normalized by the total number of grid points of the Hamiltonian
constraint for BSSN and fCCZ4 as a function of time for different values of the
parameters $\kappa_{1}$ and $\kappa_{2}$. The largest violations occur
at $t\sim 5\lambda$ when the ingoing pulse reaches the origin and
reflects (see Fig. \ref{fig:graf1}). For any value of the
damping parameters  the L2-norm of the Hamiltonian constraint is 
two orders of magnitude smaller for fCCZ4 than for BSSN at the same time. 

At times $t> 5\lambda$, different choices of the damping parameters lead
to different evolution of the L2-norm. The undamped fCCZ4 system
($\kappa_{1}=\kappa_{2}=0$) does not show any improvement in the
constraint violation with respect to BSSN.  Increasing $\kappa_{1}$
while keeping $\kappa_{2}=0$, we obtain constraint violations which
are 1 to 3 orders of magnitude smaller than with BSSN. Choosing
$\kappa_{2}=0.5$ and $\kappa_{1} \lambda =0.07$  further improves the
results. With $\kappa_{2}=-0.5$ and
$\kappa_{1} \lambda=0.07$, we find a larger violation of the constraint than
with $\kappa_{2}=\lbrace0.5,0\rbrace$. Overall, we find that these results for the
$\kappa_{2}$ parameter are similar to those reported by~\cite{Alic2}
for simulations of binary neutron stars.

\begin{figure}
\begin{minipage}{1\linewidth}
  \includegraphics[width=1.11\textwidth, height=0.3\textheight]{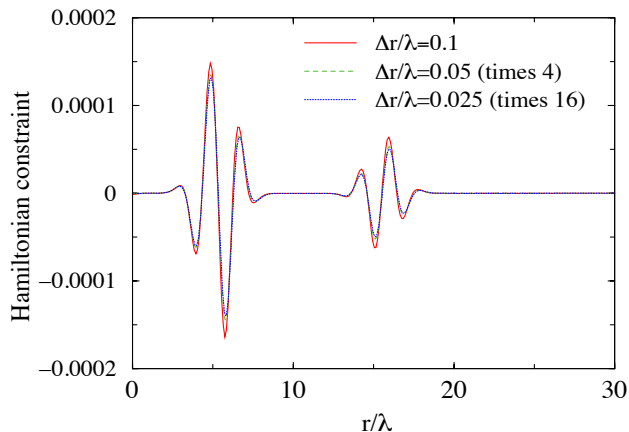}\vspace{-0.5cm}
\caption{Pure gauge pulse: Hamiltonian constraint violations at $t=10.5 \lambda$ for three different resolutions $\Delta r / \lambda=\lbrace0.1,0.05,0.025\rbrace$, rescaled by the factors corresponding to second-order convergence.}
\label{fig:graf3}
\end{minipage}
\end{figure}

We also performed three simulations with different resolutions
$\Delta r/\lambda=\lbrace0.1, 0.05, 0.025\rbrace$ to test the
convergence of the code. We show in Fig.~\ref{fig:graf3} the
rescaled Hamiltonian constraint at $t=10.5 \lambda$ for the particular choice of damping parameters
$\kappa_1 \lambda =0.07$ and $\kappa_2=0$ , 
demonstrating that the expected second-order convergence of our PIRK time-evolution
scheme is achieved.

\subsection{Schwarzschild black hole}

We next evolve a single Schwarzschild black hole given by wormhole
initial data and follow the coordinate evolution to the trumpet
geometry. We show that we are able to evolve spacetimes containing singularities without breaking the original covariance of the Z4 formulation. We use the gauge conditions
given by equations (\ref{eq14})-(\ref{eq16}), for which the evolution settles down to a maximally sliced trumpet \cite{Hannam07,BauN07}. The computational
domain has a resolution of $\Delta r=0.025 M$, $\Delta t=0.5\Delta r$ and
we use $N_{r}=60000$ grid points to place the outer boundary sufficiently 
far away from the ``puncture" at $r = 0$.

\begin{figure}
\begin{minipage}{1\linewidth}
  \includegraphics[width=1.11\textwidth, height=0.3\textheight]{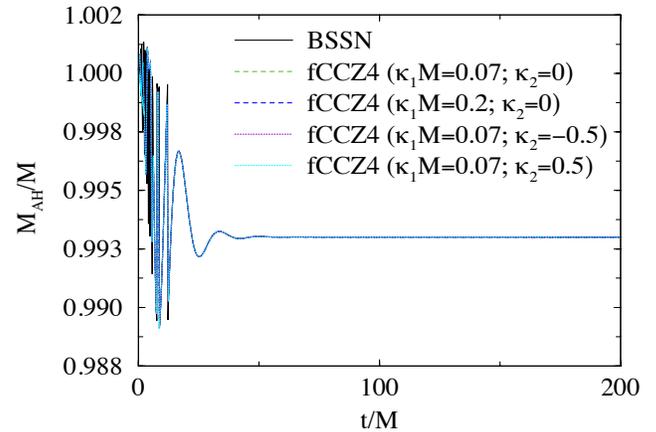}\vspace{-1cm}
\includegraphics[width=1.11\textwidth, height=0.3\textheight]{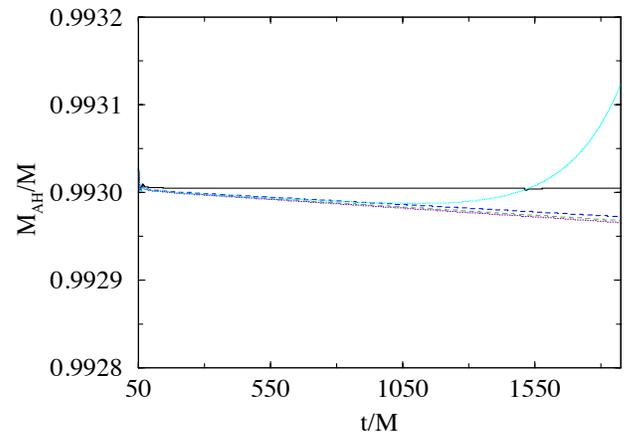}
\vspace{-0.5cm}
\caption{Time evolution of the mass of the AH in the single puncture black hole simulation. 
The lower panel shows the evolution of the AH mass during the stationary phase.}
\label{fig:graf4}
\end{minipage}
\end{figure}

\begin{figure}
\begin{minipage}{1\linewidth}
\includegraphics[width=1.11\textwidth, height=0.3\textheight]{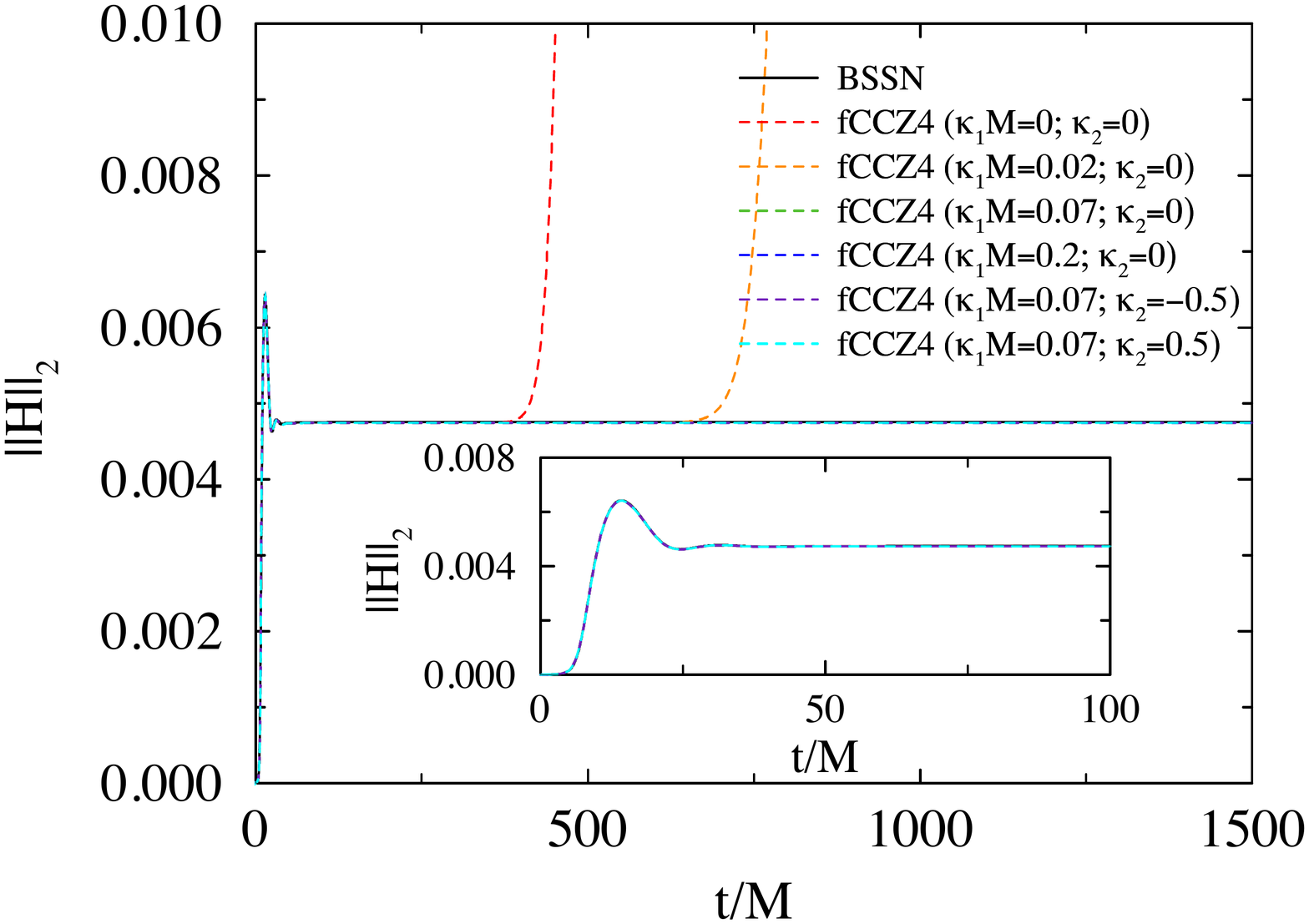}\vspace{-1cm}
\includegraphics[width=1.11\textwidth, height=0.3\textheight]{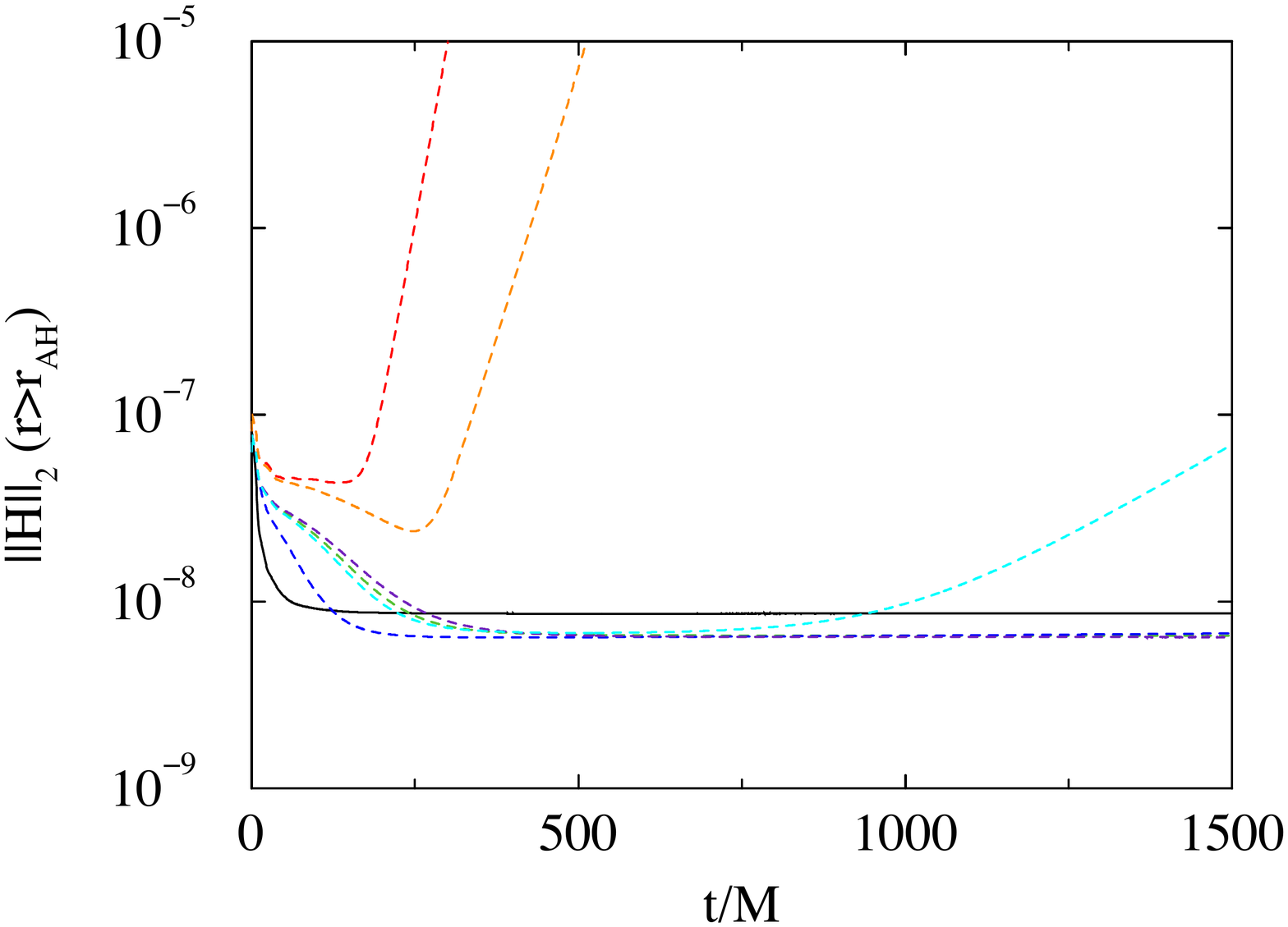}
\caption{ {\it Upper panel:} L2-norm of the Hamiltonian constraint in the single puncture black
hole simulation. The inset shows a magnified view of the initial $100M$ in the evolution. {\it Lower panel:} Same quantity but computed outside the AH.}
\label{fig:graf5}
\end{minipage}
\end{figure}

In Fig.~\ref{fig:graf4} we plot the time evolution of the apparent
horizon (AH) mass (defined as
$M_{\text{AH}}=\sqrt{\mathcal{A}/16\pi}$, where $\mathcal{A}$ is the
proper area of the horizon) for BSSN and fCCZ4. The upper panel shows this
quantity from the onset of the numerical simulation, while
the lower panel shows the AH mass only during the stationary phase
when the wormhole topology has settled to the trumpet topology. We
neither display the AH mass for the fCCZ4 system with
$\kappa_{1}=\kappa_{2}=0$ nor with $\kappa_{1} M =0.02$, $\kappa_{2}=0$
because of the appearance of numerical instabilities (see Fig.~\ref{fig:graf5}). For higher values of $\kappa_{1}$ and
$\kappa_{2}$ we obtain stable black hole evolutions.  In these cases, the difference between
the AH mass for BSSN and fCCZ4 is less than 0.005\% at the end of
the simulation ($t=1875 M$), while the error with respect to the initial ADM mass 
is $\sim 0.7$\%.   We note, however, that the black hole mass continues to 
drift for the CCZ4 formulation, while it remains constant after an initial transition for the BSSN formulation.
For the CCZ4 formulation similar results for the BH mass were 
obtained by~\cite{Alic2}, who report errors in the range 0.1-2.8\%. In contrast, the error 
is smaller for the Z4c formulation, around 0.03\% of the initial ADM
mass (see also \cite{Alic2}).   

In Fig.~\ref{fig:graf5} we plot the L2-norm of the Hamiltonian
constraint violations. The upper panel displays the L2-norm evolution in the
whole computational domain while in the lower panel we plot the
L2-norm evolution only in the region outside the AH. Clearly, the larger
violation of the Hamiltonian constraint takes place due to the
finite differencing close to the puncture, for both formulations of the
Einstein equations. However, the L2-norm of the Hamiltonian constraint
violation computed outside the AH shows that there are some
differences between the two formulations which also depend on the
values for the damping coefficients. We observe that the numerical
evolutions develop instabilities for $\kappa_{2}=0$ and
$\kappa_{1} M =(0,0.02)$. Selecting $\kappa_{1} M =0.07$ and
$\kappa_{2}=0.5$ (light blue dashed line) leads to an over-damped behavior 
that is responsible for the exponential growth of the constraint violation at late
times. We find that $\kappa_{2}=0$ with $\kappa_{1} M =0.07$ or
$\kappa_{1} M =0.2$ give the best results, leading to 
constraint violations that are comparable to those achieved with BSSN. 

Our numerical experiments with black hole initial data indicate that choosing the damping
parameter $\kappa_{2}$ different from zero does not help in reducing violations of the Hamiltonian constraint.
We therefore choose $\kappa_2 = 0$ for the remainder of the paper.

\subsection{Stable spherical relativistic star}
\label{Sec:stableTOV}

\begin{figure}
\begin{minipage}{1\linewidth}
   \includegraphics[width=1.11\textwidth, height=0.3\textheight]{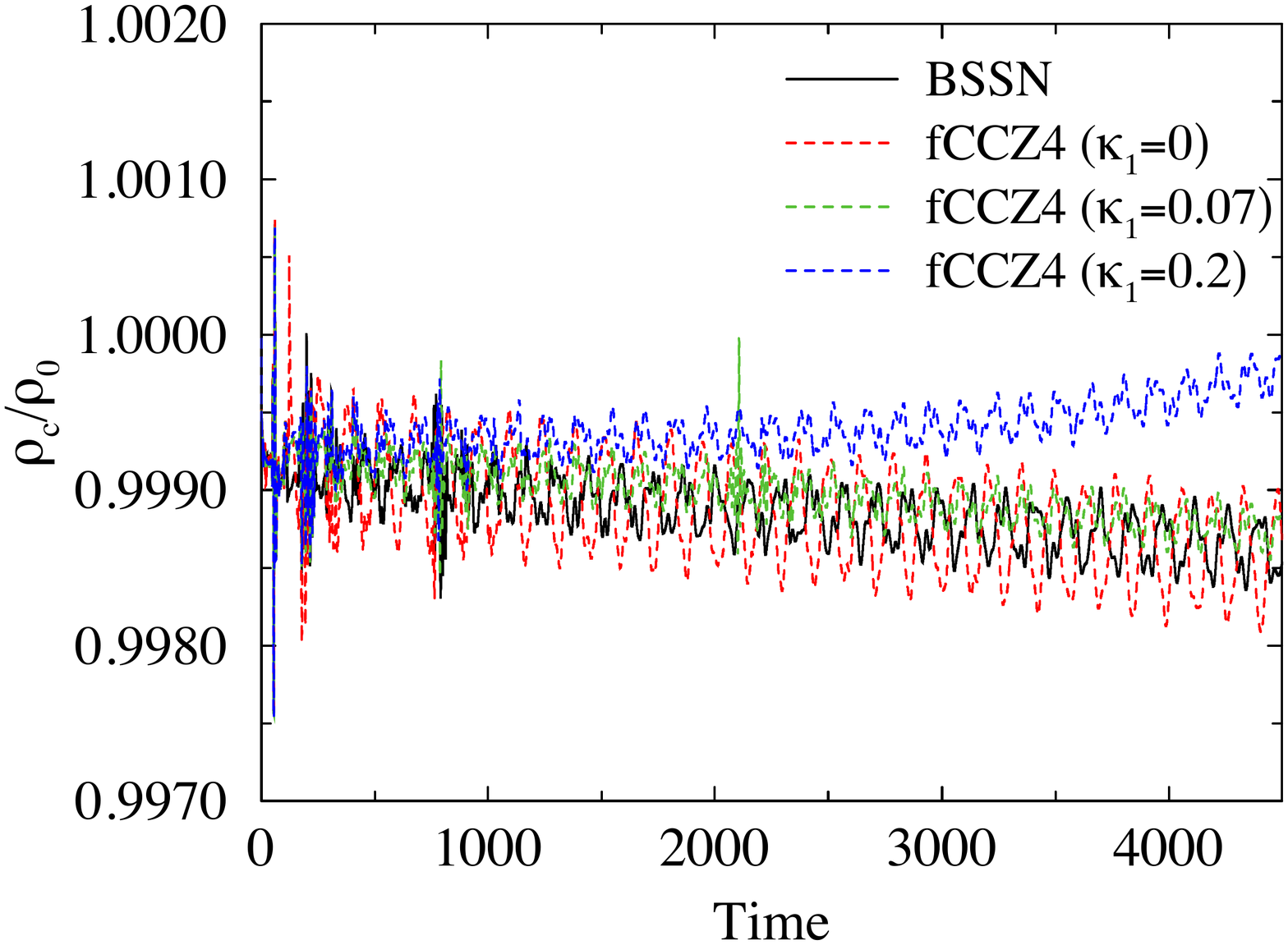}\vspace{-1cm}
\includegraphics[width=1.11\textwidth, height=0.3\textheight]{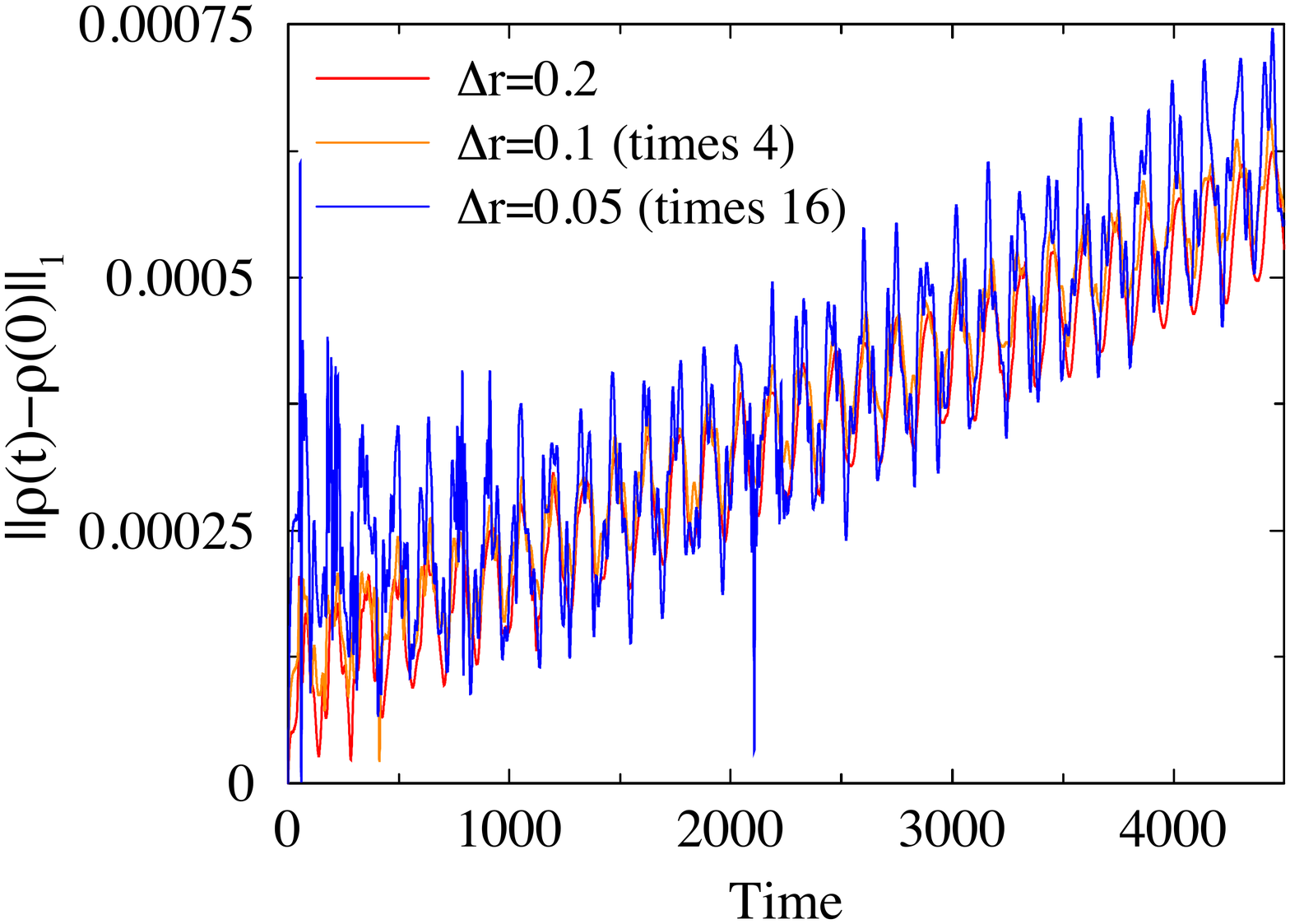}
\caption{{\it Upper panel:} Time evolution of the normalized central density with fCCZ4 for 
different values of $\kappa_{1}$ and BSSN. {\it Lower panel:} The L1 norm between  
the evolved rest-mass density and the initial density as a function of time, rescaled for three different resolutions $\Delta
r=\lbrace0.2,0.1,0.05\rbrace$ for the fCCZ4 system.}
\label{fig:graf6}
\end{minipage}
\end{figure}

In this section we turn to non-vacuum spacetimes and describe results for 
the coupled system formed by the Einstein equations and the equations of general relativistic
hydrodynamics. We construct spherically symmetric initial data by solving the Tolman-Oppenheimer-Volkoff (TOV)
equations for a polytropic equation of state
\begin{equation}
P = K \rho^{1+1/N},
\end{equation}
where $K$ is the polytropic constant and $N$ the polytropic index, and evolve these data with the Gamma-law equation of state (\ref{EOS1}) with $\Gamma = 1 + 1/N$. Throughout the remainder of the paper we will adopt $N=1$.  We also adopt code units in which $M_{\odot} = 1$; we then choose $K=100$ in these units.  In this Section we consider a star with a central density of $\rho_{c}=1.28\times10^{-3}$.  Solving the TOV equations then results in star of gravitational mass  $M=1.4M_{\odot}$, baryon rest-mass $M_{*}=1.5M_{\odot}$ and radius $R = 14.15$ km.  We evolve these initial data with $N_{r}=2000$ grid-points and a grid resolution of $\Delta r=0.05$ (so that the interior of the star is covered by approximately  200
grid-points) until a final time $t=4500$ (corresponding to 45 light crossing times).

\begin{figure}
\begin{minipage}{1\linewidth}
\includegraphics[width=1.11\textwidth, height=0.3\textheight]{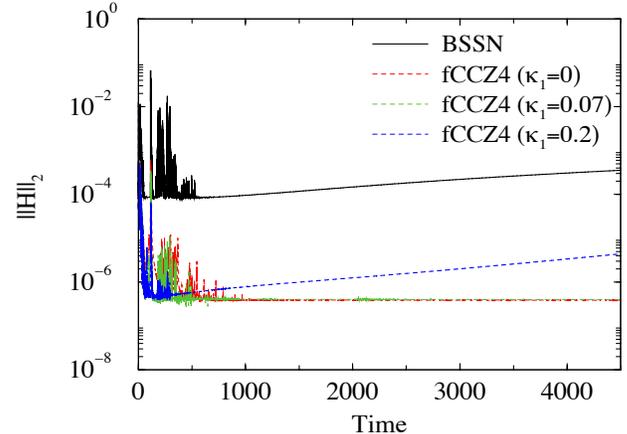}
\caption{Comparison of the time evolution of the L2-norm of the Hamiltonian constraint 
for the stable spherical relativistic star for BSSN (solid line) and fCCZ4 (dashed lines) 
with $\kappa_{1}=\lbrace0,0.07,0.2\rbrace$ (in our code units).}
\label{fig:graf7}
\end{minipage}
\end{figure}

We investigate the effect of the damping parameter $\kappa_{1}$ during the time
evolution of the TOV solution and explore the parameter space choosing
$\kappa_{1}=\lbrace0,0.07,0.2\rbrace$ in our code units, or $\kappa_1 M = \lbrace 0,  0.098, 0.28 \rbrace$. In the upper panel of Fig.~\ref{fig:graf6}, we show the 
time evolution of the normalized central rest-mass density of the star. This figure 
shows the distinctive periodic radial oscillations which are triggered by 
finite-difference errors.  These oscillations behave differently depending on 
the evolution formalism and the choices 
of the damping parameters in fCCZ4.   We find that the amplitude of 
the oscillations is reduced when the damping parameter is increased (compare  
the red dashed line and the green dashed line).  Choosing too large a value,  $\kappa_{1}=0.2$, 
causes overdamping effects which lead to a drift in the central rest-mass 
density and a growth in the L2-norm of the Hamiltonian constraint (see Fig.~\ref{fig:graf7}). 
For smaller values of $\kappa_1 M$ (i.e.~$\kappa_1 M = 0$ or 0.098)  the secular drift in the central density at late times is very similar for fCCZ4 and BSSN.   We observe that the amplitude of the oscillations decreases slightly faster for the fCCZ4 system than for BSSN, indicating that BSSN has a slightly smaller numerical viscosity.

The Fourier transform of the
time evolution of the central rest-mass density for the fCCZ4 formulation, with $\kappa_{1}M
=0.098$, agrees well with the fundamental frequency and the radial normal mode frequencies 
obtained with linear perturbation techniques \cite{Toni}. The relative 
error is less than 0.1\% for the fundamental mode and less than 0.4\% for the first three 
overtones.

We also performed a convergence test of the fCCZ4 implementation for the stable spherical star. 
In the lower panel of  Fig.~\ref{fig:graf6} we show three different curves corresponding to 
three different resolutions for the L1-norm of the difference between the evolved rest-mass density and the initial value of the density inside the star.   These findings again demonstrate second-order convergence, as expected.

Finally, in Fig.~\ref{fig:graf7} we plot the L2-norm of the Hamiltonian
constraint and we find that the constraint violations
for fCCZ4 are several orders of magnitude (at least two) smaller than for BSSN. 
The influence of the damping parameter $\kappa_{1}M$ is not
significant in the range $\lbrace0,0.098\rbrace$, but for larger values, e.g.~$\kappa_1 M=0.28$, we find an exponential growth in the L2-norm.  At a reference time $t=3000$, the L2-norm is roughly one order of magnitude larger than 
with $\kappa_1 M=\{0, 0.098\}$, but still two orders of magnitude smaller than for BSSN. 

\subsection{Migration test}
\label{Sec:migration}

Our next test of fCCZ4 is the so-called migration test of an unstable relativistic star in hydrostatic 
equilibrium~\cite{Toni}.  For this test we choose as initial data a TOV solution on the unstable branch, meaning with a density larger than that of the maximum mass configuration.  Depending on the initial perturbation, this unstable model may either collapse to a black hole, or perform initially large oscillations about a stable TOV configuration with smaller central density.   As in Section \ref{Sec:stableTOV} we adopt $N=1$ and $K=100$ (in our code units), but we now choose a central rest-mass density of $\rho_{c}=8\times10^{-3}$.  The resulting star has a gravitational mass $M=1.447$, a baryon rest-mass $M_{*}=1.535$, and a radius $R = 8.62$ km.  We evolve these data with  $N_{r}=2000$ grid-points and a resolution $\Delta r=0.025$.

\begin{figure}
\begin{minipage}{1\linewidth}
 \includegraphics[width=1.11\textwidth, height=0.3\textheight]{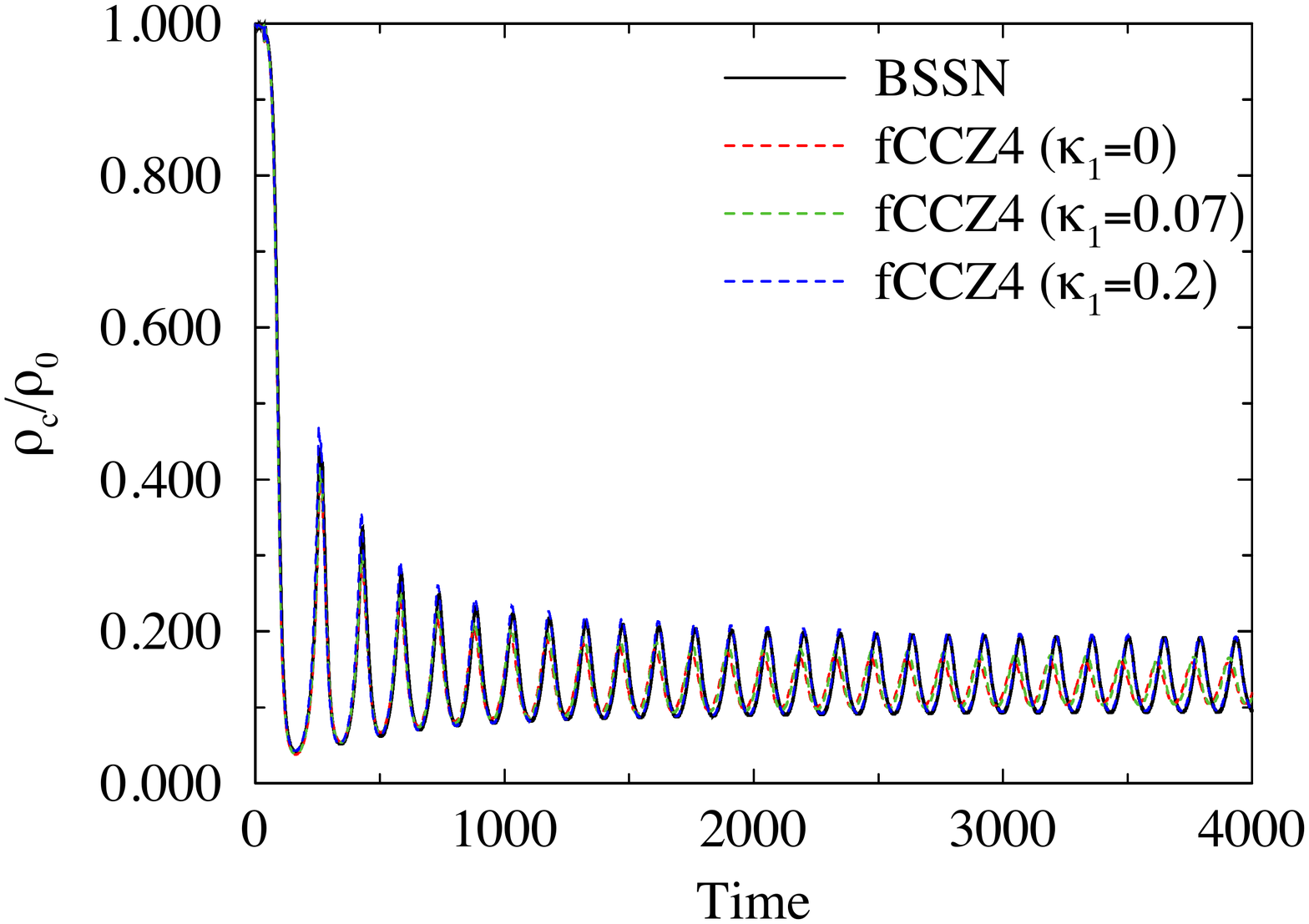}\vspace{-1cm}
\includegraphics[width=1.11\textwidth, height=0.3\textheight]{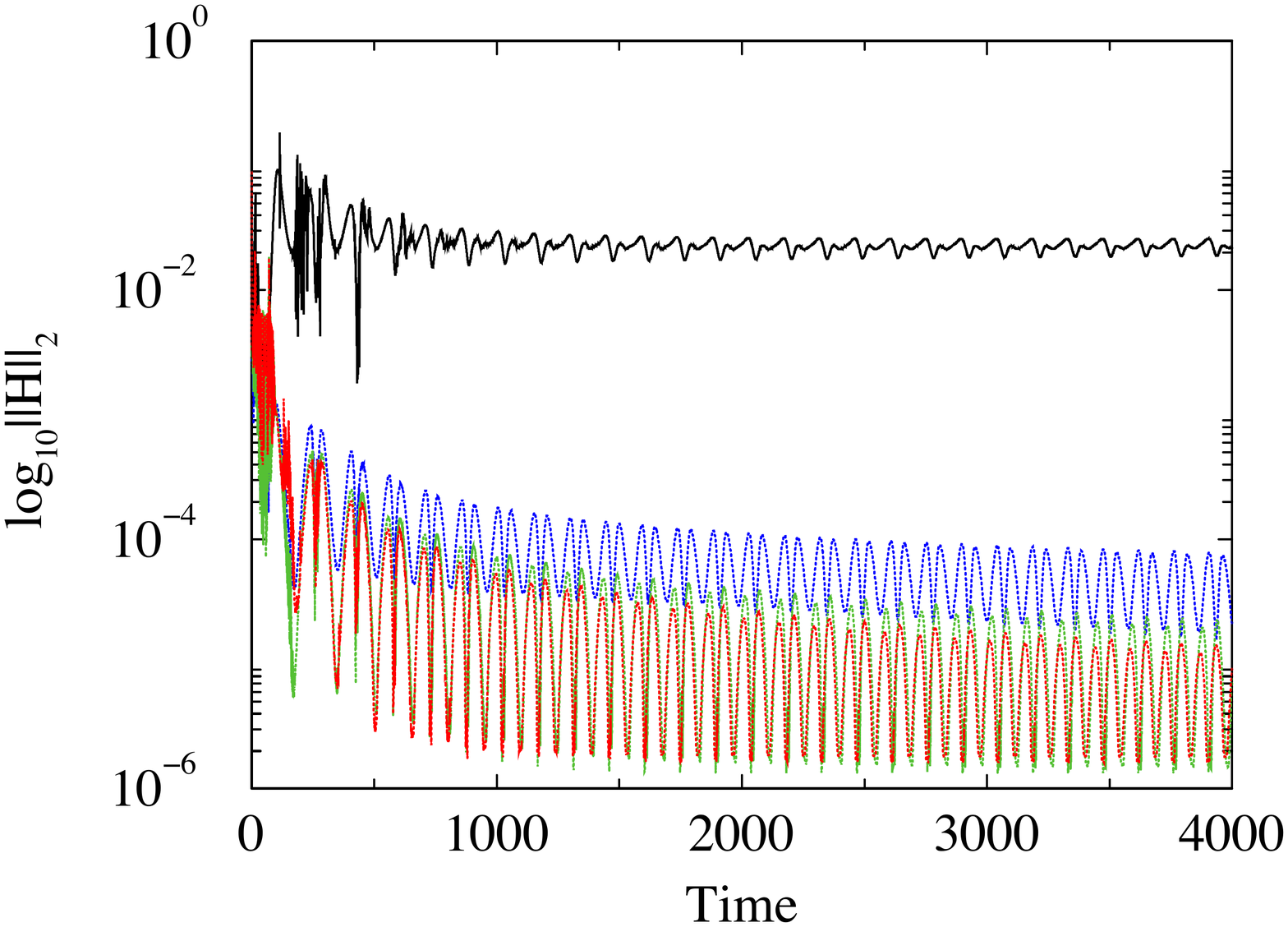}
\caption{Time evolution of the normalized central density (upper panel) and of the 
L2-norm of the Hamiltonian constraint (lower panel) for the migration test for both BSSN 
and fCCZ4 with $\kappa_{1}=\lbrace0,0.07,0.2\rbrace$ (in our code units).}
\label{fig:grafM1}
\end{minipage}
\end{figure}

In an ideal gas, the gravitational binding energy is gradually converted into internal energy via shock heating. Therefore, the high-amplitude oscillations around the new equilibrium configuration are damped and the heated stable equilibrium model approaches a central density slightly smaller than the rest-mass density of a zero temperature star of the same rest-mass ($\rho_{c}=1.35\times10^{-3}$).  This is shown in the upper panel of Fig.~\ref{fig:grafM1}, which displays the evolution of the normalized central density. After the star has migrated to the stable branch, it undergoes a series of strong expansions and contractions around the new stable configuration. During the contraction phase, shock waves are formed inside the star. When these shock waves reach the surface, small amounts of mass are expelled from the object.

Taking $\kappa_{1}=0.2$, fCCZ4 and BSSN provide very similar results
for the evolution of the central density. However, with
$\kappa_{1}=\lbrace0,0.07\rbrace$, or $\kappa_1 M = \lbrace 0, 0.10129 \rbrace$,  differences become visible at late
times. The oscillations become more damped for these values of the
damping parameter (slightly more for the undamped fCCZ4 with
$\kappa_1 M=0$), and a phase lag appears in the
oscillations. Nevertheless, the differences are not significant. The
lower panel of Fig.~\ref{fig:grafM1} shows that for the higher value
of the damping parameter, the L2 norm of the Hamiltonian constraint is
reduced by two orders of magnitude with respect to BSSN, while for the
other values of $\kappa_{1}M$ the reduction is approximately three
orders of magnitude. Another difference between BSSN and fCCZ4 is that
for the latter, the violations slightly decrease with time while they
remain constant for BSSN.  We take this as an indication that the numerical
viscosity is slightly smaller in BSSN, consistent with our findings in Section \ref{Sec:stableTOV}.
We obtain the smallest constraint violations for the smallest value of $\kappa_{1}M$, but this value also leads to the strongest damping of the oscillations.

\subsection{Gravitational collapse of a marginally stable neutron star}

The last numerical experiment is the gravitational collapse of a marginally stable spherical
relativistic star to a black hole.   As before we adopt a polytropic start with $K=100$ and $N=1$ 
as initial data, but we now consider a star with central rest-mass density $\rho_{c}=3.15\times10^{-3}$. This initial model has 
a gravitational mass
 $M=1.64$ and a baryon rest-mass $M_{*}=1.77$.   At $t=0$ we 
artificially decrease the pressure by 0.5\% in order to induce the collapse. 
We perform this test with a spatial resolution of $\Delta r=0.0125$ and $N_{r}=8000$, which places the outer boundary at $r_{\text{max}}=100$.  

\begin{figure}
\begin{minipage}{1\linewidth}
  \includegraphics[width=1.11\textwidth, height=0.3\textheight]{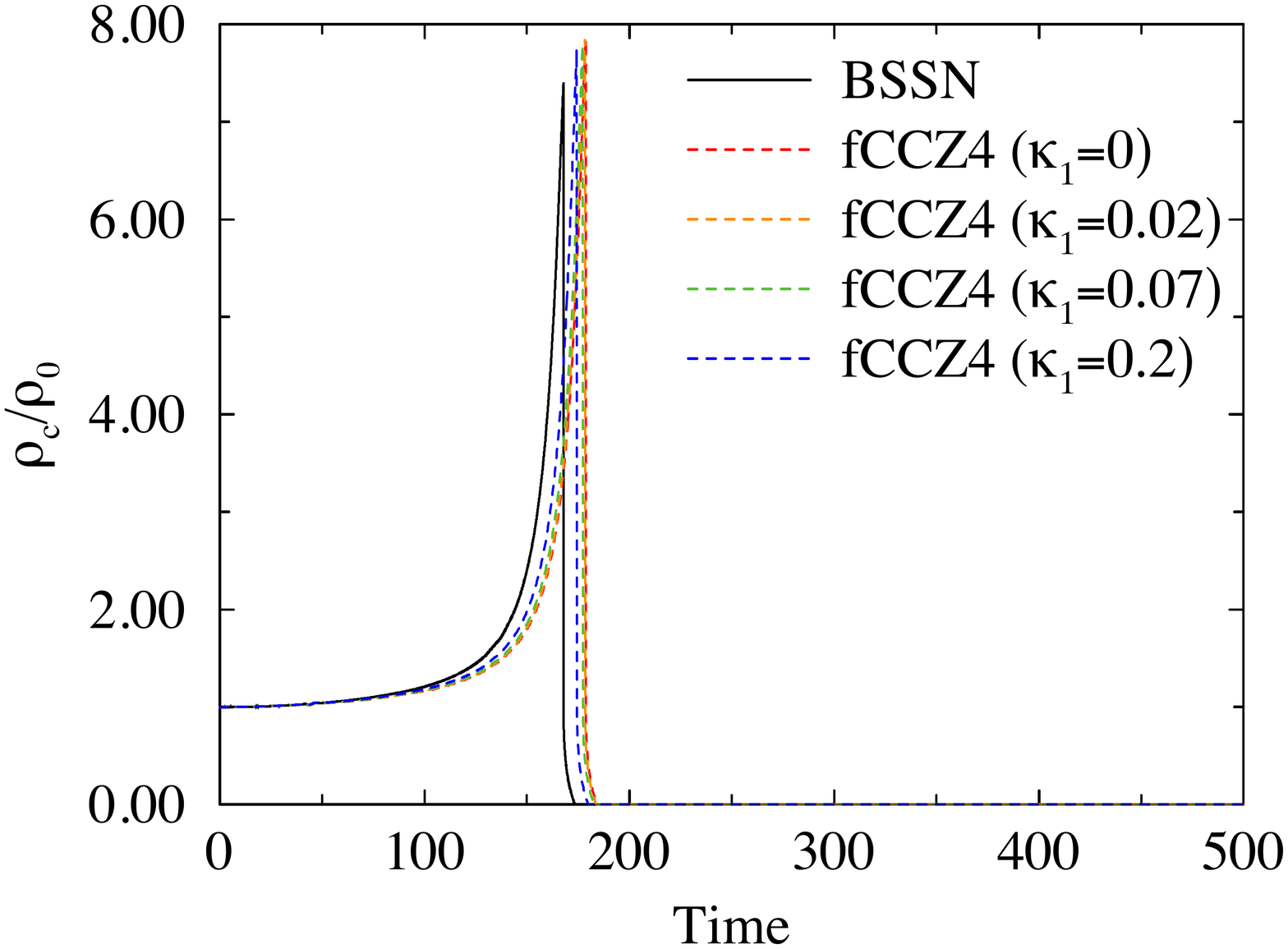}\vspace{-1cm}
\includegraphics[width=1.1\textwidth, height=0.3\textheight]{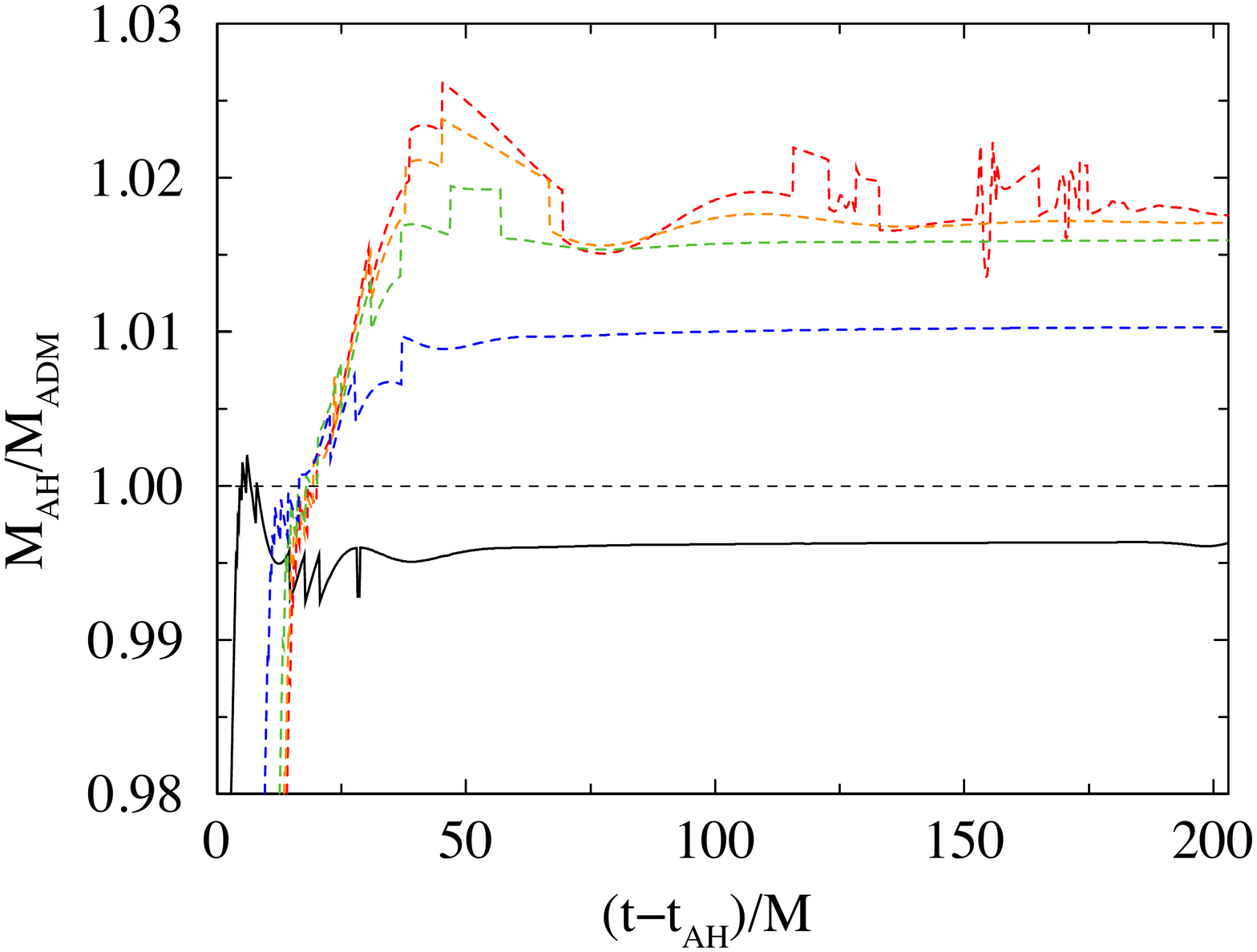}
\caption{Time evolution of the normalized central density (upper panel) and of the irreducible mass of the black hole (lower panel) for BSSN and fCCZ4 with $\kappa_{1}=\lbrace0,0.02,0.07,0.2\rbrace$ until $t=500$ (in our code units).}
\label{fig:grafC1}
\end{minipage}
\end{figure}

In Fig.~\ref{fig:grafC1}, we plot the evolution of the normalized central density and the 
mass of the AH after it forms at a time $t_{\rm AH}$.  We find that $t_{\rm AH}$ depends slightly on the formulation used, and, for fCCZ4, on the coefficient $\kappa_1 M$: for BSSN, we found $t_{\rm AH} \sim 167$, while for fCCZ4 with $\kappa_1 M = 0$ we found $t_{\rm AH} \sim 177$ (all in our code units).   Increasing $\kappa_1 M$ slightly reduces $t_{\rm AH}$, as shown in Fig.~\ref{fig:grafC1}.  This behavior is again consistent with our observations in Section \ref{Sec:stableTOV} and \ref{Sec:migration}, and suggests that the numerical viscosity of the BSSN scheme is slightly smaller than that of fCCZ4. It also suggests that the numerical viscosity of fCCZ4 decreases with increasing $\kappa_1 M$.   For  $\kappa_{1}M=0.82$ (not shown in Fig.~\ref{fig:grafC1}), $t_{\rm AH}$ agrees well  with that of BSSN, although this choice of $\kappa_{1}M$  leads to non-negligible over-damped results (an important drift for the black hole mass appears).  
 
In the lower panel of Fig.~\ref{fig:grafC1} we show the horizon mass as a function of time, as obtained with the different evolution schemes.  The difference between the  initial ADM  mass of the system
 and the mass of the AH at $t=500$ for the BSSN formulation is about
 0.5\%.  We find a slightly higher deviation, around 1.6-1.7\% for fCCZ4 with $\kappa_{1}M=\lbrace0.0,0.1148\rbrace$ and 1\% for $\kappa_{1}M=0.328$.

\begin{figure}
\begin{minipage}{1\linewidth}
\hspace{-0.5cm}\includegraphics[width=1.11\textwidth, height=0.3\textheight]{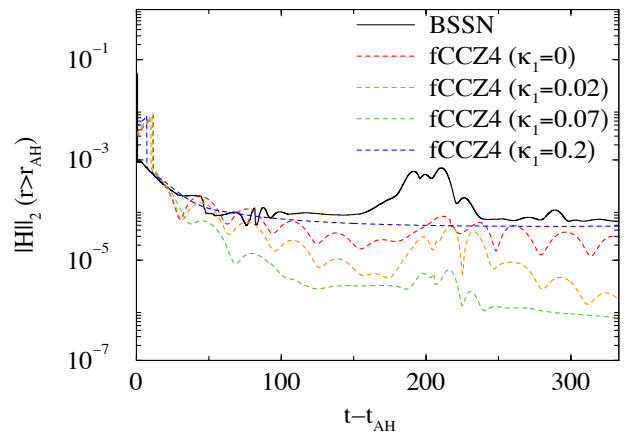}
\caption{Time evolution of the L2-norm of the Hamiltonian constraint
  for BSSN and fCCZ4 computed outside of the AH. The time coordinate
  is relative to the time $t_{\text{AH}}\sim167$ when an apparent horizon forms for the BSSN formulation.}
\label{fig:grafC2}
\end{minipage}
\end{figure}

Finally, in Fig.~\ref{fig:grafC2} we show the L2-norm of the Hamiltonian
constraint violation computed in the region outside the AH. As
expected, the constraint violations obtained with the fCCZ4 formulation are
smaller than those obtained with the BSSN formulation. The difference in the constraint
violations between the two formulations is at most three orders of
magnitude when we take $\kappa_{1} M=0.1148$ (green dashed line in 
Fig.~\ref{fig:grafC2}) . Increasing the value of $\kappa_{1}$ does 
not reduce the L2-norm further.  Instead it increases again and approaches a value 
similar to that obtained with BSSN.

\section{Summary}
\label{section:summary}

In this paper we generalize the covariant and conformal Z4 system \cite{Bona} of the Einstein equations originally proposed by Alic et al.~\cite{Alic1} using a reference-metric
approach~\cite{FCF} (see \cite{Gourgoulhon07,Brown,Gourgoulhon12} for the derivation of the BSSN system using the same approach).   The resulting system, which we refer to as fCCZ4,
allows us to write the evolution equations in a fully covariant form suitable for curvilinear coordinate systems.  As a first step, we implement the fCCZ4 system in spherical coordinates 
under the assumption of spherical symmetry.  We adopt a PIRK scheme for the time evolution and obtain stable evolutions -- without regularization of the equations -- for both vacuum and non-vacuum spacetimes. 

The CC4Z formalism of \cite{Alic1,Alic2} shares some properties with a similar approach, Z4c,  developed by \cite{Sebastiano1}.   In agreement with \cite{Alic1,Alic2}, we find that using Sommerfeld
outer boundary conditions is as accurate as it is for the BSSN
system.  Unlike in the Z4c formalism, we therefore find stable evolutions even without implementing
constraint preserving boundary conditions~\cite{Ruiz}.  Unlike reported in \cite{Alic1}, we did not need to introduce a third parameter $\kappa_3$ in order to obtain stable evolutions for black-hole spacetimes (see \cite{Alic2} for an alternative modification of the equations).

We performed a number of tests to compare the accuracy of the fCCZ4
formulation with that of the BSSN system. As in previous experiments with Z4c and CCZ4, we find that constraint violations in neutron-star spacetimes are significantly smaller in fCCZ4 than in BSSN, often by 2 or 3 orders of magnitude.  We find similar improvements for the pure gauge-wave test.  This test also demonstrates that fCCZ4 can reduce errors introduced by the coordinate singularities in spherical polar coordinates, even though this effect was less visible in our other simulations.  We note, however, that these results depend on the choices for the free parameters $\kappa_1$ and $\kappa_2$.  Poor choices may introduce over-damping, thereby increasing errors, or may make the code unstable.   We also note that our findings suggest that the fCCZ4 scheme introduces a slightly larger numerical viscosity than the BSSN scheme.

For black-hole spacetimes, we find that, even for the best choices for the free parameters, fCCZ4 reduces the constraint violations only very moderately, and only at late times.  At least for the resolutions that we employed in our tests, BSSN was slightly more accurate in computing the black hole mass (compare \cite{Alic1,Sebastiano1}).

We found that choosing $\kappa_{2}$ different from zero did not lead to significant improvements; in fact, poor choices may lead to over-damping effects.  On the other hand, the damping parameter
$\kappa_{1}$ plays an important role in the reduction of the violations of the constraints.  Increasing the value of this parameter tends to reduce constraint violations (except in the migration test) but may also introduce a damping that is too large, thereby making the code unstable and causing it to crash. For all examples considered in this paper we have been able to find suitable choices for $\kappa_1$, but for more general applications it may be difficult to identify an optimal choice for this parameter.   Since $\kappa_1$ has units of inverse length, its optimal choice depends on typical length-scales in the simulation.  For a single black hole, for example, a good choice appears to be $\kappa_1 \simeq 0.07/M$.   In simulations of black hole binaries with unequal masses, on the other hand, it may be hard to find a single parameter $\kappa_1$ that is well-suited for both black holes.  Similar issues may arise in other mixed systems, e.g.~black hole-neutron star binaries or black holes surrounded by accretion disks.   An optimal choice of $\kappa_1$ for the matter component, for example, might lead to over-damping for the black hole.  Should this issue indeed prove to be a problem, a possible solution might be to allow $\kappa_1$ to take different values in different regions of the spacetime.

In a future project we will implement the fCCZ4 formalism in three spatial dimensions without any symmetry assumptions, and we plan to explore the issues discussed above with that code. 


\acknowledgments

NSG thanks the Max-Planck-Institut
f\"ur Astrophysik for its hospitality during the development of part
of this project. PM thanks Sebastiano Bernuzzi and David Hilditch for
helpful discussions.  TWB gratefully acknowledges support from the Alexander-von-Humboldt Foundation.  This work was supported in part by the Spanish
MICINN (AYA 2010-21097-C03-01), by the Generalitat Valenciana (PROMETEO-2009-103), by the Deutsche Forschungsgesellschaft (DFG) through 
its Transregional Center SFB/TR 7 ``Gravitational Wave Astronomy'', and by NSF grant PHY-1063240 to Bowdoin College.

\appendix
\section{Detailed source terms included in the PIRK operators for the evolution
equations}
\label{appendix}
The evolution Eqs.~(\ref{eq28})-(\ref{eq30}), (\ref{eq33}), (\ref{eq34}),
(\ref{eq36}), (\ref{eq39}) (\ref{eq14})-(\ref{eq16}), are evolved using a second-order PIRK
method, described in Sec.~III. In this Appendix we provide a complete listing of the source terms 
included in the explicit or partially implicit operators.

Firstly, the hydrodynamic conserved quantities and the spacetime fields 
$a$, $b$, $X$, $\alpha$ and 
$\beta^r$, are evolved explicitly, i.e., all the source terms of the evolution 
equations of these variables are included in the $L_1$ operator of the 
second-order PIRK method.

Secondly, $A_a$, $K$ and $\Theta$ are evolved partially implicitly, using updated values 
of $\alpha$, $a$ and $b$; more specifically, the corresponding $L_2$ and $L_3$ 
operators associated with the evolution equations for $A_a$, $K$ and $\Theta$ are:
\begin{align}
	L_{2(A_a)} &= - \left(\nabla^{r}\nabla_{r}\alpha 
- \frac{1}{3}\nabla^{2}\alpha\right) 
+ \alpha\left(R^{r}_{r} - \frac{1}{3}R\right) \nonumber\\
&+ \alpha\bigl(\mathcal{D}^{r}Z_{r}+\mathcal{D}_{r}Z^{r}-\frac{2}{3}\mathcal{D}_{m}Z^{m}\bigl),\\
	L_{3(A_a)} &= \alpha (K-2\Theta) A_{a} - 16\pi\alpha(S_a - S_b)\nonumber\\
&+\beta^{r}\partial_{r}A_{a}, \\
	L_{2(K)} &= - \mathcal{D}^{2}\alpha+\alpha\bigl(R+2\mathcal{D}_{m}Z^{m}\bigl), \\
	L_{3(K)} &= \beta^{r} \partial_{r}K  
+ \alpha(K^{2}-2\Theta K) -3\alpha\kappa_{1}(1+\kappa_{2})\Theta\nonumber \\
& + 4\pi\alpha(S_{a} + 2S_{b}-3E),\\
L_{2(\Theta)} &= - Z^{r}\partial_{r}\alpha+\frac{1}{2}\alpha\bigl(R+2\mathcal{D}_{m}Z^{m}\bigl), \\
	L_{3(\Theta)} &= \beta^{r} \partial_{r}\Theta  
+ \frac{1}{2}\alpha(A_{a}^{2} + 2A_{b}^{2} + \frac{2}{3}K^{2}-2\Theta K) \nonumber \\
& -\alpha\kappa_{1}(2+\kappa_{2})\Theta- 8\pi\alpha E.
\end{align}

Next, $\tilde{\Lambda}^{r}$ is evolved partially implicitly, using updated values 
of $\alpha$, $a$, $b$, $\beta^r$, $\phi$, $A_a$, $K$ and $\Theta$; more 
specifically, the corresponding $L_2$ and $L_3$ operators associated with the 
evolution equation for $\tilde{\Lambda}^{r}$ are:
\begin{align}
	L_{2(\tilde{\Lambda}^{r})} &= \frac{1}{a}\partial^{2}_{r}\beta^{r} 
+ \frac{2}{b}\partial_{r}\left(\frac{\beta^r}{r}\right)
+ \frac{\sigma}{3 a}\partial_{r}(\hat{\nabla}_m\beta^{m}) \nonumber \\
  & - \frac{2}{a}(A_{a}\partial_{r}\alpha + \alpha\partial_{r}A_{a}) 
- \frac{4\alpha}{r b}(A_{a}-A_{b}) \nonumber \\
	& + \frac{\xi \alpha}{a} \left[\partial_{r}A_{a} 
- \frac{2}{3}\partial_{r}K + 6A_{a}\partial_{r}\chi  \right. \nonumber \\
  & \left. + (A_{a}-A_{b})\left(\frac{2}{r} + \frac{\partial_{r}b}{b}\right)
\right] + 2\alpha A_{a}\bar{\Lambda}^{r}\nonumber\\
&- \bar{\Lambda}^{r}\partial_{r}\beta^{r}
+ \frac{2\sigma}{3}\bar{\Lambda}^{r}\hat{\nabla}_m\beta^{m} \nonumber \\
&+\frac{2}{a}\biggl(\alpha\partial_{r}\Theta-\Theta\partial_{r}\alpha-\frac{2}{3}\alpha KZ_{r}\biggl)\nonumber\\
&\frac{2}{a}\biggl(\frac{2}{3}Z_{r}\mathcal{\bar D}_{m}\beta^{m}-Z_{r}\partial_{r}\beta^{r}\biggl)-\frac{2}{a}\kappa_{1}Z_{r}, \\
	L_{3(\tilde{\Lambda}^{r})} &= \beta^{r}\partial_{r}\tilde{\Lambda}^{r}
- 8\pi j_{r} \frac{\xi \alpha}{a}.
\end{align}

Finally, $B^r$ is evolved partially implicitly, using updated values of 
$\tilde{\Lambda}^{r}$, i.e., 
$\displaystyle L_{2(B^r)} = \frac{3}{4}\partial_{t}\tilde{\Lambda}^{r}$ and 
$L_{3(B^r)} = 0$.



\end{document}